\documentclass[reprint, amsmath,amssymb, aps,superscriptaddress]{revtex4-1}

\usepackage{graphics}
\usepackage{dcolumn}
\usepackage{bm}
\usepackage[caption=false]{subfig}
\usepackage{xfrac}
\usepackage{mathrsfs}
\usepackage{float}
\usepackage[normalem]{ulem}
\usepackage[bbgreekl]{mathbbol}
\usepackage[utf8]{inputenc}
\usepackage{amsmath}
\usepackage{multirow}
\usepackage{adjustbox}

\newcommand{\figref}[1]{Fig.~\ref{fig:#1}}
\newcommand{\tabref}[1]{Table.~\ref{table:#1}}
\newcommand{\secref}[1]{Sec.~\ref{#1}}
\newcommand{\secsref}[2]{Sec.~\ref{#1}--\ref{#2}}
\renewcommand{\eqref}[1]{(\ref{eq:#1})}
\newcommand{\eqsref}[2]{(\ref{eq:#1},\ref{eq:#2})}
\newcommand{\eqssref}[2]{(\ref{eq:#1}-\ref{eq:#2})}
\renewcommand{\vec}[1]{\mathbf{#1}}

\newcommand{\mat}[1]{\mathbb{#1}}

\newcommand{\appendixsec}[2]{\section{\uppercase{#1}}\label{#2}}

\newcommand{\appendixTRY}[3]{\section{\uppercase{#1}{#2}}\label{#3}}

\newcommand{\Scale}[2][4]{\scalebox{#1}{$#2$}}%

\begin{document}

\title{Ab-initio multimode linewidth theory for arbitrary inhomogeneous laser cavities}

\author{A. Pick}
\affiliation{Department of Physics, Harvard University, Cambridge, Massachusetts 02138, USA}
\author{A. Cerjan}
\affiliation{Department of Applied Physics, Yale University, New Haven, Connecticut 06520, USA}
\author{D. Liu}
\affiliation{Department of Physics, Massachusetts Institute of Technology, Cambridge, Massachusetts 02139, USA}
\author{A. W.  Rodriguez}
\affiliation{Department of Electrical Engineering, Princeton University, Princeton, New Jersey 08544, USA}
\author{A. D. Stone}
\affiliation{Department of Applied Physics, Yale University, New Haven, Connecticut 06520, USA}
\author{Y. D. Chong}
\affiliation{Division of Physics and Applied Physics, School of Physical and Mathematical Sciences,
Nanyang Technological University, Singapore 637371, Singapore}
\author{S. G. Johnson}
\affiliation{Department of Mathematics, Massachusetts Institute of Technology, Cambridge, Massachusetts 02139, USA}

\begin{abstract}
We present a multimode laser-linewidth theory for arbitrary cavity structures and geometries that contains nearly all previously known effects and also finds new nonlinear and multimode corrections, 
e.g. a correction to the  $\alpha$ factor due to openness of the cavity and a multimode Schawlow--Townes relation (each linewidth is proportional to a sum of inverse powers of all lasing modes).
Our  theory produces a  quantitatively accurate formula for the linewidth, with no free parameters, including the full spatial degrees of freedom of the system. Starting with the Maxwell--Bloch equations, we handle quantum and thermal noise by introducing random currents whose correlations are given by the fluctuation--dissipation theorem. We derive coupled-mode equations for the lasing-mode amplitudes and obtain a formula for the linewidths in terms of simple integrals over the steady-state lasing modes.  
\end{abstract}

\maketitle

\section{Introduction \label{intro}}

The fundamental limit on the linewidth of a laser is a foundational question in laser theory~\cite{Sargent1974,Haken1984,Haken1985,Svelto1976,Milonni2010}. It  arises from quantum and thermal fluctuations~\cite{Gordon1955,Schawlow1958}, and depends on many parameters of the laser (materials, geometry, losses, pumping, etc.); it remains an open problem to obtain a fully general linewidth theory. In this paper, we present a multimode laser-linewidth theory  for arbitrary cavity structures and geometries that contains nearly all previously known effects~\cite{Petermann1979,Lax1966,Exter1995,Henry1982,Osinski1987} and also finds new nonlinear and multimode corrections. The theory is quantitative and makes no significant approximations; it simplifies, in the appropriate limits, to the Schawlow--Townes  formula \eqref{STP} with the well-known corrections.  It also demonstrates the interconnected behavior of these corrections~\cite{Chong2012,Pillay2014}, which are usually treated as independent. Most previous laser-linewidth theories have employed simple models for calculating the lasing modes (e.g., making the paraxial approximation).  Such simplifications, though appropriate for many macroscopic lasers, are inadequate for describing complex microcavity lasers such as 3d nanophotonic structures or random lasers with inhomogeneities on the wavelength scale~\cite{He2013,Painter1999,Loncar1999,Park2004}.  We base our theory on the recent steady-state ab-initio laser theory (SALT)~\cite{Tureci2006,Ge2010}, which allows us to  efficiently  solve the semi-classical laser equations in the absence of noise for arbitrary structures~\cite{Esterhazy2013}. 
We treat the noise as a small perturbation to the SALT solutions, allowing us to  obtain  the linewidths \emph{analytically}  in terms of simple integrals over the  steady-state lasing modes. Our SALT-based theory is {\it ab initio}  in the sense that it produces  quantitatively accurate formulas for the linewidths, with no free parameters, including the full spatial degrees of freedom of the system. 
 Hence, we will refer to this approach as the noisy steady-state \emph{ab-initio} laser theory (N-SALT).

Our derivation (Secs.~III--V) begins with the Maxwell--Bloch equations (details in appendix A), which couple the full-vector Maxwell equations to an atomic gain medium~\cite{Lamb1964}, combined with random currents (in \secref{FDTsec}) whose statistics are described by the fluctuation--dissipation theorem (FDT)~\cite{Callen1951,Rytov1989,Dzyaloshinkii1961,Lifshitz1980,Eckhardt1982}. In the presence of these random currents, the  amplitudes  of the lasing modes evolve  according to a set of coupled ordinary differential equations (ODEs), which have been called ``oscillator models''~\cite{Lax1967a,Henry1986} or ``temporal coupled-mode theory'' (TCMT)~\cite{Haus1984,Haus1991,Suh2004,Jannopoulos2008,Rodriguez2007} in similar contexts.  In their most general form, our N-SALT TCMT equations (\secref{CMT}) have the form of oscillator equations with a \emph{non-instantaneous} nonlinear term that stabilizes the mode amplitudes around their steady-state values. The non-instantaneous nonlinearity arises since the atomic populations respond with a time delay to field fluctuations; this corresponds to the typical case of ``class B" lasers~\cite{Arecchi1984,Oppo1986,Lugiato1984}, in which the population dynamics cannot be adiabatically eliminated.  We are able to show analytically that the resulting linewidths of the lasing peaks are identical to the results one obtains for a simplified model with instantaneous nonlinearity~\cite{Lax1967a,Henry1986}, which describes the (less common) case of ``class A" lasers, in which the population dynamics are adiabatically eliminated. As expected, however, in certain parameter regimes the full non-instantaneous model can exhibit side peaks alongside the main lasing peaks~\cite{Exter1992}, arising from relaxation oscillations(\secref{spectrum}.C).

By solving  the N-SALT TCMT equations, we obtain a simple closed-form matrix expression for the linewidths and multimode phase correlations (\secref{spectrum}), generalizing earlier  two-mode results that used phenomenological models~\cite{Elsasser1985}.  This gives a multimode ``Schawlow--Townes'' relation (\secref{applications}.C), where the linewidth of each lasing mode is proportional to a sum of inverse output powers of the neighboring lasing modes. 
The theory is valid well above threshold, and whenever a new mode turns on, this inverse-power relation produces a divergence  due to the failure of the linearization approximation near threshold. However, we show that this divergence is spurious and can be avoided by solving the nonlinear N-SALT TCMT equations numerically~\cite{Hui1993}. (Our formalism can be extended to treat the near-threshold regime analytically by including noise from sub-threshold modes, as discussed in \secref{applications}.B and in \secref{summary}.)  \secsref{applications}{3d} also present several other model calculations that illustrate the differences between N-SALT and previous linewidth theories. Finally, in \secref{summary}, we discuss some potential additional corrections that will be addressed in future work.  
In a second manuscript~\cite{Cerjan2014}, we also compare the theory against full time-dependent integration of the stochastic Maxwell--Bloch equations and find excellent quantitative agreement with the major results presented here.
\begin{figure}
                 \includegraphics[width=0.4\textwidth]{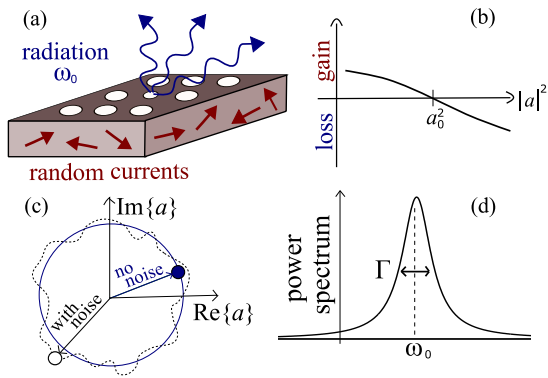}
		\caption{(Color online) Schematics illustrating linewidth physics. (a)~Photonic-crystal (PhC) laser cavity~\cite{Jannopoulos2008} emitting radiation from the lasing mode at frequency~$\omega_0$, perturbed by random currents.  
(b)~The squared amplitude is stabilized around~$a_0^2$.  Below (above)~$a_0^2$, the medium provides light amplification (attenuation). 
(c)~Phasor diagram for the complex field amplitude: a circular oscillation (with~$|a|=a_0$) for the noise-free mode and a perturbed path for noise-driven mode. Noise drives small amplitude fluctuations and possibly large phase drifts.
(d)~The lineshape is a Lorentzian~$\sim \Gamma/[(\omega-\omega_0)^2 + (\frac{1}{2}\Gamma)^2]$, centered around~$\omega_0$ with width~$\Gamma$.}
\label{fig:illustration}
\end{figure}

Laser dynamics are surveyed in many sources~\cite{Sargent1974,Haken1984,Haken1985,Svelto1976,Milonni2010}, but it is useful to review here a simple physical picture of linewidth physics. A resonant cavity [e.g., light bouncing between two mirrors or a photonic-crystal (PhC) microcavity as in \figref{illustration}(a)] traps light for a long time in some volume, and lasing occurs when a gain medium is ``pumped'' to a population ``inversion'' of excited states to the point (\emph{threshold}) where gain balances loss.  [Of course, this simple picture is modified once additional modes reach threshold, or for lasers (such as random lasers~\cite{Wiersma2008,Tureci2008}) in which the passive cavity possesses no strong resonances; all of these complexities are handled by SALT~\cite{Tureci2006,Ge2010} and hence are incorporated into our approach.]   For simplicity, consider here a laser operating in the single-mode regime. Above threshold, the gain depends nonlinearly on the mode intensity~$|a|^2$, as sketched in \figref{illustration}(b): increasing the field intensity decreases the gain due to depletion of the excited states  until it reaches a  stable steady-state value~$a_0^2$. (This gain-saturation effect is called ``spatial hole-burning''~\cite{Svelto1976} since it can be spatially inhomogeneous.) In the absence of noise, this results in a stable sinusoidal oscillation with an infinitesimal linewidth, but the presence of noise, which can be modeled by random current  fluctuations~$\vec{J}$~\cite{Henry1986,Duan1990,Exter1995}, perturbs the mode as depicted in \figref{illustration}(c), resulting in a finite linewidth. There are various sources of noise in real lasers, but spontaneous emission sets a fundamental lower limit on the linewidth~\cite{Svelto1976}; here we will include only spontaneous emission and thermal noise.  In particular, although the squared  amplitude is stabilized around $a_0^2$ by the nonlinear gain, the phase~$\phi$ of the mode drifts according to a random walk (a Brownian/Wiener phase) with variance~$\langle \phi^2 \rangle \approx \Gamma\hspace{2pt} t$, and the Fourier transform of a Wiener phase yields a Lorentzian lineshape 
[\figref{illustration}(d)] with full width at half maximum (FWHM)~$\Gamma$~\cite{Lax1967a}. The goal of linewidth theory is to derive~$\Gamma$, ideally given only the thermodynamic FDT description of the current fluctuations and the Maxwell--Bloch physics of the laser cavity.

The most basic approximation for the linewidth (sufficiently far above threshold), usually referred to as the Schawlow-Townes (ST) formula~\cite{Gordon1955,Schawlow1958}, takes the form 
\begin{equation}
\Gamma = \frac{\hbar\omega_0\gamma_0^2}{2P},
\label{eq:ST}
\end{equation}
where~$P$ is the output power of the laser,~$\gamma_0$ is the passive cavity resonance width, and~$\omega_0$ is the laser frequency, often approximated to be equal to the real part of the passive-cavity resonance pole at~$\omega_*=\omega_0-i\gamma_0/2$. 
(A slightly more accurate approximation for the laser frequency takes into account the small line-pulling of the laser frequency towards the atomic transition frequency~\cite{Siegman1986}.) The inverse-power dependence causes the famous line-narrowing of a laser above threshold.

Over the decades, a number of now-standard corrections to this formula were found~\cite{Haken1985,Svelto1976,Milonni2010}, leading to  the modified ST formula:
\begin{equation}
\Gamma = \frac{\hbar\omega_0{{\gamma}_0}^2}{2P}\cdot
n_\mathrm{sp}\cdot
{\left|\frac{\displaystyle\int_\mathrm{C} dx \hspace{2pt}{|\vec{E}_\mathrm{c}|}^2}{\displaystyle\int_\mathrm{C} dx\hspace{2pt} \vec{E}_\mathrm{c}^2}\right|}^2\cdot
\left(\frac{\gamma_\perp}{\gamma_\perp+\frac{\gamma_0}{2}}\right)^2
\cdot
\left(1+\alpha_0^2\right).
\label{eq:STP}
\end{equation}
First, the gain medium can be thought of, in many respects, as a system at negative temperature~$T$~\cite{Patra2000}, with the limit of complete inversion of the two lasing levels corresponding to~$T \to 0^-$.  When only partial inversion is present, the linewidth is enhanced by a factor of $n_\mathrm{sp}\equiv\frac{N_2}{N_2-N_1}$~\cite{Kuppens1994,Kuppens1996}, where $N_2$ and $N_1$ are the \emph{spatially averaged} populations  in the upper and lower states of the lasing transition. We refer to this correction as the incomplete-inversion factor (also known as  ``the spontaneous emission factor'').
Second, due to the openness of the laser system, the modes are not power-orthogonal and the noise power which goes into each lasing mode is enhanced~\cite{Siegman1989b}; this correction is known as the Petermann factor, and it becomes significant in low-$Q$ laser systems, where it is not a good approximation to treat the lasing mode $\vec{E}_\mathrm{c}$ as purely real. ($Q\equiv\omega_0/\gamma_0$ is a dimensionless passive-cavity lifetime defined in units of the optical period~\cite{Jannopoulos2008}.) 
Note that $\vec{E}_\mathrm{c}$ is the passive-cavity mode [in contrast to SALT solutions, which are the modes of the full non-linear equations, introduced in \eqref{SALTsols}].
  $\int_\mathrm{C} dx$ denotes integration over the cavity region.  
Third, for low-$Q$ laser cavities, it is possible that the gain linewidth $\gamma_\perp$ can be on the order of or smaller than the passive cavity resonance width $\gamma_0$, causing significant dispersion effects as the gain is increased to threshold~\cite{Lax1966}.  This correction is commonly called  the ``bad-cavity" factor~\cite{Kuppens1994B,Exter1995}.  Unlike the other corrections mentioned above, the bad-cavity factor decreases the laser linewidth. However, very few lasers systems are in the parameter regime where this effect is significant~\cite{Kuppens1995}. Finally, amplitude fluctuations in the laser field couple to the phase dynamics, leading to a correction known as the ``$\alpha$ factor''.  For atomic gain media, this effect was identified by Lax \cite{Lax1966} in the 1960's, and for this case it is typically a small correction.  For bulk semiconductor gain media the effect is large, and typically dominates the broadening due to direct phase fluctuations~\cite{Vahala1983,Westbrook1987,Lang1986}; in this context it is known as the ``Henry $\alpha$ factor'' \cite{Henry1982}.

Previous linewidth derivations have taken a number of different approaches, making severe approximations compared to the solution of the full three-dimensional space-dependent Maxwell--Bloch equations in the presence of noise.  Generally speaking, linewidth theories can be classified into two categories.  The first class includes methods which solve Maxwell's equations with a phenomenological model for the gain medium and account for noise spatial and spectral correlations by using the FDT~\cite{Henry1986,Duan1990,Exter1995}. Typically, these methods do  not handle nonlinear spatial hole-burning above threshold or multimode effects.  These methods, commonly used in the semiconductor laser literature, resulted in linewidth formulas which included the Petermann~\cite{Siegman1989b}, bad-cavity~\cite{Haken1984,Exter1995}, incomplete-inversion~\cite{Henry1986}, and $\alpha$ factors~\cite{Henry1982}. 
Most notably, an early work by Arnaud~\cite{Arnaud1986} derived a single-mode linewidth formula without making any simplifying assumptions about the field patterns,  handling anisotropic, inhomogeneous, and dispersive media. However, this theory was only applied to  very simple, effectively one-dimensional,  homogeneous systems, and it  was missing hole-burning effects and the $\alpha$ factor.

The second class of linewidth theories  consists of scattering-matrix methods~\cite{Schomerus2000,Schomerus2009,Chong2012,Pillay2014}, which can 
treat arbitrary geometries without phenomenological parameters and take into account the effects of spatial hole-burning.  S-matrix theories only have access to the input and output fields and, therefore, can only treat the noise in a spatially averaged manner and are not able to obtain the $\alpha$ factor  rigorously. However,  they obtain all of the other corrections to the single-mode linewidth. In particular, the recent S-matrix approach by Chong  \emph{et al.}~\cite{Chong2012,Pillay2014} takes advantage, as we do, of the \emph{ab-initio}  computational approach of SALT, and hence has the potential to treat arbitrary geometries and  spatial hole-burning effects.  
 (We reduce  our results to the most recent scattering-matrix linewidth formula~\cite{Pillay2014} in appendix D.) Note that in practice, S-matrix methods require a substantial independent calculation beyond SALT to extract the linewidths, whereas our approach obtains the linewidths immediately from  SALT calculations (or any other method to obtain the steady-state lasing modes) by simple integrals over the fields.

Our derivation of N-SALT, being based on the SALT solutions, has a similar regime of validity.
For \emph{single-mode lasing}, SALT and N-SALT are essentially exact, relying only on the rotating-wave approximation and on the laser being sufficiently far above threshold.  For multimode lasing, those theories require two additional dynamical constraints~\cite{Tureci2006,Ge2010}: the rates associated with population dynamics must be small compared to both the dephasing rate of the polarization and the lasing mode spacing (roughly, the free spectral range).  The former constraint is satisfied in all solid-state lasers, whereas the latter requires a sufficiently small laser cavity.  The actual size depends both on details of the cavity and of the gain medium used, but the appropriate limit is realized in many complex lasers of interest. When these frequency scales are not well-separated, the level populations are not quasi-stationary, and multimode SALT will initially lose accuracy and eventually fail completely (since multimode lasing becomes  unstable~\cite{Ge2008}). Moreover, while the average (SALT) behavior is unaffected by non-lasing poles, they do affect the noise properties, and N-SALT in its current form only accounts for a finite number of poles in the Green's function (appendix A.2). [We only include lasing poles (i.e., poles on the real axis), but extension to include non-lasing poles, which determine the amplified spontaneous emission (ASE)~\cite{Hui1993,Siegman1989a},  will be straightforward (\secref{summary})].  As noted above, the linewidth formula additionally assumes that the laser  is operating far enough above threshold that  amplitude fluctuations are small compared to the steady state amplitudes (i.e.,~$|a(t)|\approx a_0$ in the notation of \secref{spectrum}). Hence, our formula does not describe the linewidth near the lasing thresholds. Our perturbation approach takes into account only the lowest-order correction to the complex modal amplitude~$a(t)$ and neglects higher-order corrections to the  frequency~$\omega_0$ and spatial pattern~$\vec{E}_0(\vec{x})$ [see Eq.~\eqref{pertAnsatz}].  Moreover, we neglect non-Lorentzian corrections to the lineshape~\cite{Scully1988a,Scully1988b,Benkert1990a, Benkert1990b,Kolobov1993} (\secref{FDTsec}). In the following section we present our generalized  linewidth formula in the single-mode regime \eqref{TCMTwidth} and compare it with traditional linewidth theories.

\section{The N-SALT linewidth formula \label{formula}}

Our main result is a multimode linewidth formula which generalizes \eqref{STP}.  In the multimode case, the result takes the form of a covariance matrix for the phases of the various modes, which is presented in \eqsref{multiAlpha}{multiWidth} of \secref{spectrum}.  In the single-mode case, the N-SALT linewidth formula takes the simple form:
\begin{align}
\Gamma = \frac{\hbar\omega_0{\widetilde{\gamma}}_0^2}{2P}
\cdot\widetilde{n}_\mathrm{sp}\cdot\widetilde{K}\cdot\widetilde{B}
\cdot(1+\widetilde{\alpha}^2).
\label{eq:TCMTwidth}
\end{align}
The modified  correction  factors (marked by tildes) are defined in \tabref{Factors}.  As can be seen from the table, those factors generalize the traditional expressions  by taking into account  both spatial inhomogeneity and nonlinearity. Since  the generalized factors depend on the SALT permittivity $\varepsilon$, mode profile $\vec{E}_0(\bold{x})$, and frequency $\omega_0$, one can no longer regard the effects of  cavity-openness, nonlinearity, and dispersion  as separate multiplicative effects. In this sense, our formula demonstrates the intermingled nature of the linewidth correction factors, as previously introduced in \cite{Chong2012,Pillay2014}, but here demonstrated in a new level of generality.
We denote by $\int dx$ integration over  all space, for any number of spatial dimensions. We use the shorthand notation for vector products $|\vec{E}_0|^2=\vec{E}_0\cdot\vec{E}_0^*$ and $\vec{E}_0^2=\vec{E}_0\cdot\vec{E}_0$, 
where the latter unconjugated inner product appears naturally because of the biorthogonality relation for lossy complex-symmetric systems~\cite{Moiseyev2011,Siegman2000}. 
$\mbox{Im}\hspace{2pt}\varepsilon (\vec{x})$ denotes the imaginary part of the nonlinear steady-state  permittivity \eqref{epsiBODY}, which is negative/positive in gain/loss regions. 
The  output power $P$ is related to the SALT solutions  by invoking Poynting's theorem, which one can use to show that $P \propto \int_\mathrm{P} dx\hspace{2pt}[-\mbox{Im}\hspace{2pt}\varepsilon(\vec{x})]|\vec{E}_0(\vec{x})|^2$. We use $\int_\mathrm{P} dx$ to denote some volume which contains the gain medium. The choice of the volume is somewhat arbitrary; e.g., integrating over the cavity region corresponds to the output power at the cavity boundary~\cite{Henry1986}. Note, however, that this arbitrariness in the choice of the volume is not a general feature of our formula. After substituting the relevant expressions from \tabref{Factors} into \eqref{TCMTwidth}, the integrals which contain $\int_\mathrm{P} dx$    cancel, resulting in an expression for the linewidth only in terms of integrals over the entire space.
The effective inverse temperature $\beta(\vec{x})$ is determined by the inhomogeneous steady-state atomic populations $N_1(\vec{x})$ and $N_2(\vec{x})$, and is defined as~\cite{Jeffers1993,Matloob1997,Patra1999}
\begin{equation}
\beta(\vec{x})\equiv  \frac{1}{\hbar\omega_0}\ln\left(\frac{N_1(\vec{x})}{N_2(\vec{x})}\right).
\label{eq:Beta}
\end{equation}
In regions where the gain medium is pumped sufficiently to invert the population, $\beta(\vec{x})$ is negative; in  regions where the pump is too weak to invert, $\beta(\vec{x})$ will be positive [and still given by \eqref{Beta}]; and in unpumped regions, Eq. \eqref{Beta} will simply reduce to the equilibrium temperature of the surrounding environment ${(k_\mathrm{B}T)}^{-1}$. The quantities $N_1(\vec{x})$ and $N_2(\vec{x})$ are an output of the SALT solution in the absence of noise. The spatially dependent expression inside the square brackets in the definition of $\widetilde{n}_\mathrm{sp}$  in \tabref{Factors} generalizes the spatially averaged incomplete-inversion factor $\frac{N_2}{N_2-N_1}$. That can be seen by noting that  $\frac{1}{2}\coth(\frac{\hbar\omega\beta}{2})-\frac{1}{2}=(\exp[\hbar\omega\beta]-1)^{-1}\equiv n_\mathrm{B}$, where $n_\mathrm{B}$ is the  usual Bose--Einstein distribution function~\cite{Landau1980,Kittel1980}. (For gain media, it is sometimes convenient to introduce the positive spontaneous-emission factor $n_\mathrm{sp}=-n_\mathrm{B}$~\cite{Henry1996}. Note that this definition ensures that the generalized incomplete-inversion factor is always positive.)
The $\frac{1}{2}$ factor subtracted from the hyperbolic cotangent  was discussed in~\cite{Henry1996}, and we give a simple classical explanation for it in appendix E. If standard absorbing layers are used to implement outgoing boundary conditions in the SALT solver~\cite{Esterhazy2013} and the temperature of the ambient medium is assigned to these layers, then the N-SALT formula  includes the effect of incoming thermal radiation.
A generalized Petermann factor which formally resembles $\widetilde{K}$ appeared in previous work by Schomerus \cite{Schomerus2009} (in his expression for the Petermann factor for TM modes in two-dimensional dielectric resonators). However, the earlier formula is expressed in terms of  passive resonance scalar fields, whereas our correction contains  3d nonlinear SALT solutions.  Finally,~$\widetilde{\alpha}$ is a generalized $\alpha$ factor, defined explicitly in \secref{spectrum} \eqref{alpha}.   For atomic gain media, the traditional factor is expressed in terms of  the atomic transition frequency $\omega_\mathrm{a}$ and decay rate of the atomic polarization $\gamma_\perp$. In the current work we will only evaluate the atomic case, although the general expression in terms of the non-linear coupling $C$ should also apply to the semiconductor case.

\begin{table}
\setlength{\tabcolsep}{0.08in}	
{\footnotesize
\begin{tabular}{c c c}
Symbol  & Traditional & Generalized \\[0.1ex] \hline\hline \\[1pt]
\begin{tabular}{@{}c@{}c@{}} $\widetilde{\gamma}_0$ \\ cavity decay\\rate\end{tabular}
 &$\gamma_0$& 
$\left|\frac{\displaystyle\int dx    \,(\omega_0\mbox{Im}\hspace{2pt}\varepsilon) {\vec{E}_0}^2}
{\displaystyle\int dx  \hspace{2pt}\varepsilon\, {\vec{E}_0}^2}\right|$\\[7ex]
  \begin{tabular}{@{}c@{}c@{}}  $\widetilde{n}_\mathrm{sp}$ \\ incomplete \\inversion\end{tabular} 
  & $\displaystyle\frac{N_2}{N_2-N_1}$&
  $\Scale[0.9]{\frac{\displaystyle\int\!\!dx    \Scale[0.9]{\left[\tfrac{1}{2}\coth(\tfrac{\hbar\omega\beta}{2})\!-\!\tfrac{1}{2}\right]}\mbox{Im}\varepsilon {|\vec{E}_0|}^2}
{\displaystyle\int_\mathrm{P} dx  \,\mbox{Im}\hspace{2pt}\varepsilon \,{|\vec{E}_0|}^2}}$ \\[7ex]
\begin{tabular}{c@{}c@{}}  $\widetilde{K}$ \\ Petermann\end{tabular}& 
  $\Scale[0.9]{{\left|\tfrac{\displaystyle\int_\mathrm{C} dx \hspace{2pt}{|\vec{E}_\mathrm{c}|}^2}{\displaystyle\int_\mathrm{C} dx\hspace{2pt} \vec{E}_\mathrm{c}^2}\right|}^2}$&
  $\Scale[0.95]{\left|\tfrac{\displaystyle\int_\mathrm{P} dx    \,\mbox{Im}\hspace{2pt}\varepsilon\, {|\vec{E}_0|}^2}
{\displaystyle\int dx  \,\mbox{Im}\hspace{2pt}\varepsilon\, {\vec{E}_0}^2}\right|^2}$  \\[7ex]
\begin{tabular}{c@{}c@{}}  $\widetilde{B}$  \\ bad cavity\end{tabular}
& $\left(\frac{\gamma_\perp}{\gamma_\perp+\frac{\gamma_0}{2}}\right)^2$&
$\Scale[0.9]{\displaystyle\left|\tfrac{\displaystyle\int dx\, \varepsilon\,\vec{E}_0^2}{{\displaystyle\int dx \hspace{2pt}\vec{E}_0^2 \left(\varepsilon+\tfrac{\omega_0}{2}\tfrac{\partial\varepsilon}{\partial\omega_0}\right)}}\right|^2}$ \\[7ex]
\begin{tabular}{c@{}c@{}}  $\widetilde{\alpha}$  \\ amplitude-phase \\ coupling\end{tabular}
 & $\frac{\omega_\mathrm{a}-\omega_0}{\gamma_\perp}$ &
$\frac{\mbox{Im}\hspace{2pt}C}{\mbox{Re}\hspace{2pt}C}$ \\[7ex]
\begin{tabular}{@{}c@{}c@{}} $C$  \\ nonlinear\\ coupling\end{tabular} &   &
$\Scale[0.8]{\tfrac{-i\tfrac{\omega_0}{2}
\displaystyle\int dx \tfrac{\partial\varepsilon}{\partial |a|^2} \vec{E}_0^2}
{\displaystyle\int dx \left(\varepsilon+\tfrac{\omega_0}{2}\tfrac{\partial\varepsilon}{\partial\omega_0}\right)\vec{E}_0^2}}$ 
\end{tabular}}
\caption{Traditional and new linewidth correction factors for the single-mode linewidth formulas \eqsref{STP}{TCMTwidth}.}
\label{table:Factors}
\end{table}

The N-SALT  formula \eqref{TCMTwidth} reduces to the traditional formula \eqref{STP} in some limiting  cases. 
Let us consider, for simplicity, a 1d  Fabry-P{\'e}rot laser cavity of length $L$ surrounded  by  air (i.e., $\mbox{Im}\hspace{2pt}\varepsilon=0$ outside the cavity region). Let us assume also that the laser is operating not too far above the threshold and is uniformly pumped, hence $\mbox{Im}\hspace{2pt}\varepsilon$ and 
$\beta$ are nearly constant inside the cavity. In this limit, all the integrals in \tabref{Factors} can be approximated by reducing the integration  limits to the cavity region; terms which contain integration over the imaginary part of the permittivity are  non-zero only within the cavity region (e.g., $\int dx\hspace{2pt}\mbox{Im}\hspace{2pt}\varepsilon|\vec{E}_0|^2$ becomes  $\mbox{Im}\hspace{2pt}\varepsilon\int_\mathrm{C} dx\hspace{2pt}|\vec{E}_0|^2$); while terms of the form 
$\int dx\, \varepsilon\,\vec{E}_0^2$ can be written as the sum of the cavity contribution $\varepsilon\int_\mathrm{C} dx\hspace{2pt}\vec{E}_0^2$ and the surrounding medium contribution $\int_{\mbox{out}}dx \hspace{2pt}\vec{E}_0^2$, where the latter  is negligible for  $L\omega_0\gg1$, as shown in  appendix D and in~\cite{Pillay2014} (here and throughout the paper, we are setting $c=1$). 
Using this approximation, it is immediately apparent from \tabref{Factors} that the incomplete-inversion factor reduces to the traditional expression. The generalized Petermann factor reduces to the traditional factor in the limit of a high-Q cavity, where the threshold lasing state $\vec{E}(\vec{x})$ is approximately the same as the passive resonance state $\vec{E}_\mathrm{c}(\vec{x})$. In order to simplify the remaining terms, recall that the lasing threshold is reached when gain in the system compensates for the loss.  For weak losses (small $\mbox{Im}\hspace{2pt}\varepsilon/\varepsilon$) that can be treated by perturbation theory, the threshold condition is  $\gamma_0=\frac{\omega_0\mbox{Im}\hspace{2pt}\varepsilon}{\varepsilon}$~\cite{Haken1984} and, therefore, the generalized decay rate reduces to $\gamma_0$  (one can thereby see that the Schawlow--Townes formula \eqref{STP} neglects nonlinear corrections to $\gamma_0$, as was also shown in~\cite{Chong2012}). Next, let us discuss the generalized bad-cavity factor, which  simplifies to 
$\left(1+\frac{\omega_0}{2\varepsilon}\frac{\partial\varepsilon}{\partial\omega_0}\right)^{-2}$ after reducing  the integration limits. In order to show that it agrees with the traditional  factor, we need to show that  $\frac{\omega_0}{2\varepsilon}\frac{\partial\varepsilon}{\partial\omega_0}\approx \frac{\gamma_0}{2\gamma_\perp}$.
The steady-state  effective  permittivity, as used in SALT theory (appendix A.1), is
\begin{equation}
\varepsilon(\vec{x})=
\varepsilon_\mathrm{c}(\vec{x}) + \frac{\gamma_\perp D(\vec{x})}{\omega_0-\omega_\mathrm{a}+i\gamma_\perp},
\label{eq:epsiBODY}
\end{equation}
where $\varepsilon_\mathrm{c}$ is the passive permittivity and the second term is the active nonlinear permittivity due to the gain medium. The population inversion   $ D(\vec{x})  = N_2 (\vec{x})-N_1(\vec{x})$ is generally spatially varying above threshold due to spatial hole-burning. 
Since we assume here that we are close to threshold and that the pumping is
uniform,  the inversion is also  uniform in space and near its threshold value. If one assumes, additionally, that the detuning of the lasing frequency from atomic resonance is small ($|\omega_0-\omega_\mathrm{a}|\ll\gamma_\perp$), one obtains 
$\frac{\partial\varepsilon}{\partial\omega_0}\approx\frac{\mbox{Im}\hspace{2pt}\varepsilon}{\gamma_\perp}$.   Finally, we show in \secref{applications}.A that our $\widetilde{\alpha}$ reduces to the known $\alpha_0$ in homogeneous low-loss cavities, so that all factors of the corrected ST formula
are recovered in this limit. (Note that line-pulling effects which may modify the lasing frequency $\omega_0$ are handled by SALT.)

In the next section, we  present the TCMT equations which  are used  in this paper to derive the N-SALT linewidth  formula \eqref{TCMTwidth}, but which may also  be used  to extract more information on laser dynamics away from steady state.

\section{The N-SALT TCMT  equations \label{CMT}}

In the absence of noise, the electric field of a  laser operating in the multimode regime is given by 
 the real part of $\vec{E}_0(\vec{x},t)$, where  
\begin{equation}
\vec{E}_0(\vec{x},t)=\sum_\mu \vec{E}_\mu(\vec{x})a_{\mu 0}e^{-i\omega_\mu t},
\label{eq:SALTsols}
\end{equation}
and the laser has zero linewidth. 
(This assumes, of course, that there exists a  steady-state multimode solution of the nonlinear semi-classical lasing equations~\cite{Tureci2006,Ge2010}.)
The modes $\vec{E}_\mu(\vec{x})$ and frequencies $\omega_\mu$ can be  calculated 
using SALT, which solves the semi-classical Maxwell-Bloch equations in the absence of noise. (SALT has been generalized to include multi-level atoms~\cite{Cerjan2012}, multiple lasing transitions, and gain diffusion~\cite{Cerjan2014b}; any of these cases can thus be treated by N-SALT with minor modifications, but we focus on the two-level case here.)  The linewidth can now be calculated by adding Langevin noise, as described below. 

In the presence of a weak noise source, the electric field can be written as a superposition of the steady-state lasing modes with time-dependent amplitudes $a_\mu(t)$ which fluctuate around $a_{\mu 0}$:
\begin{equation}
\vec{E}(\vec{x},t)=\sum_\mu \vec{E}_\mu(\vec{x})a_\mu(t)e^{-i\omega_\mu t}.
\label{eq:pertAnsatz}
\end{equation}
In principle, the sum in \eqref{pertAnsatz} should also include the non-lasing modes since the set of lasing modes by itself does not form a complete basis for the fields. Non-lasing modes contribute to amplified spontaneous emission (ASE), which has a significant effect on the  spectrum near and below the lasing thresholds~\cite{Hui1993,Siegman1989a} and will be treated in future work.

In appendix A, we derive the N-SALT TCMT equations of motion for $a_\mu (t)$ starting with  the full vectorial Maxwell-Bloch equations.
We show that the noise-driven field obeys an effective nonlinear equation which,  in the frequency domain, takes the form
\begin{equation}
\left[\nabla\times\nabla\times-\omega^2
\varepsilon(\omega,a)\right]\widehat{\vec{E}}(\vec{x},\omega)
=\widehat{\vec{F}}_\mathrm{S}(\vec{x},\omega),
\label{eq:ME-Body}
\end{equation}
where the carets denote Fourier transforms~[e.g.,~$\vec{E}(\vec{x},t)\equiv \int_0^\infty d\omega\,e^{-i\omega t}\widehat{\vec{E}}(\vec{x},\omega)$]. Spontaneous emission is included via the stochastic noise term~$\widehat{\vec{F}}_\mathrm{S}(\vec{x},\omega)$ (quantified in \secref{FDTsec}), and the effective permittivity $\varepsilon(\omega,a)$ (derived in appendix A.2) is given by
\begin{equation}
\Scale[0.95]{
\varepsilon(\omega,a) \widehat{\vec{E}}(\vec{x},\omega)=
\sum_\mu\left[\varepsilon_\mathrm{c} \widehat{a}_\mu+\frac{\gamma_\perp}{\omega-\omega_\mathrm{a}+i\gamma_\perp}\widehat{D}*\widehat{a}_\mu\right]
\vec{E}_\mu(\vec{x})},
\label{eq:epsiBody}
\end{equation}
where the asterisk denotes a convolution. The second argument of $\varepsilon(\omega,a)$ denotes the implicit  dependence of $\varepsilon$ on the modal amplitudes  $a_\mu$  through the inversion $\widehat D$. The effective permittivity \eqref{epsiBody} can be decomposed into  a steady-state-amplitude dispersive term and a nonlinear non-dispersive term (similar in spirit to~\cite{Dana2014}). The key point here is that, to lowest order, there are two corrections to the permittivity in the presence of noise: the \emph{dispersive correction} due to any shift in frequency at the unperturbed amplitudes $a_{\mu0}$, and the \emph{nonlinear correction} due to any shift in amplitude at the unperturbed frequency. (Shifts in frequency are small because only frequency components within the mode linewidths matter, while shifts in amplitude are small because of the stabilizing effect of gain feedback.) The coupling between these two perturbations is higher order and is hence dropped, which greatly simplifies the analysis.
 
Substituting the permittivity expansion (derived explicitly in appendix A.3) into Maxwell's equation \eqref{ME-Body}, we find that the noise-driven field obeys the linearized  equation
\begin{equation}
\Scale[0.98]{
\left[\nabla\times\nabla\times-\omega^2
\varepsilon(\omega,a_0)\right]\widehat{\vec{E}}(\vec{x},\omega)
=\widehat{\vec{F}}_\mathrm{NL}(\vec{x},\omega)
+\widehat{\vec{F}}_\mathrm{S}(\vec{x},\omega)},
\label{eq:ME-effBody}
\end{equation}
i.e., the dispersive permittivity which appears on the left-hand side of \eqref{ME-effBody} is evaluated at the steady-state amplitude $a_0$. The nonlinear non-dispersive term $\widehat{\vec{F}}_\mathrm{NL}$ [defined explicitly in \eqref{FNL}], which corresponds to amplitude fluctuations at the unperturbed frequency, appears as a restoring force on the right-hand side. The noise-driven field $\widehat{\vec{E}}(\vec{x},\omega)$ is found in appendix A.4 by convolving the linearized Green's function with the source terms $\widehat{\vec{F}}_\mathrm{NL}$ and $\widehat{\vec{F}}_{S}$. Finally, the N-SALT TCMT equations are obtained by transforming the noise-driven field back into the time domain. 

\subsection{Time-delayed multimode model}

We find that,  in the most general case, the  TCMT equations take the form
\begin{align}
\dot{a}_\mu&=
 \sum_\nu 
\int dx \, c_{\mu\nu}(\vec{x})\,\times
\nonumber\\
      &\left[
 \gamma(\vec{x})\int^t dt'  e^{-\gamma(\vec{x})(t-t')}
\left(a_{\nu0}^2-|a_{\nu}(t')|^2\right) 
\right]a_\mu+f_\mu.
\label{eq:timeDelay}
\end{align}
Comparing \eqref{timeDelay} and \eqref{ME-effBody}, one can see that 
the first term on the right-hand side of \eqref{timeDelay} is related to the  nonlinear restoring force $\widehat{\vec{F}}_\mathrm{NL}$,
and the  Langevin noise  $f_\mu (t)$ is associated with $\widehat{\vec{F}}_{S}$.

The nonlinear coupling coefficients $c_{\mu\nu}(\vec{x})$   [derived in \eqref{Cderiv}] correspond to local changes in the nonlinear permittivity with respect to intensity changes in each of the modes
\begin{equation}
c_{\mu\nu}=\frac{-i \omega_\mu^2\frac{\partial\varepsilon(\omega_\mu)}{\partial |a_\nu|^2} \vec{E}_\mu^2}
{\displaystyle\int dx(\omega_\mu^2\varepsilon)'_\mu \vec{E}_\mu^2},
\label{eq:coupling}
\end{equation}
where we have introduced a shorthand notation for the derivative in the denominator 
   $ (\omega_{\mu}^2\varepsilon)'_{\mu}\equiv \left.\frac{\partial }{\partial\omega} \omega^2\varepsilon \right|_{\omega_{\mu}}$.
This modal coupling in the fluctuation dynamics 
comes about because of saturation of the gain:  a fluctuation in mode $\mu$ affects the amplitudes of 
all the other modes $\nu$.
%

The N-SALT TCMT equations are nonlocal in time because the atomic populations are not in general able
to follow the field fluctuations instantaneously and, instead, respond with a time delay determined
by the local atomic decay rate $\gamma(\vec{x})$, given by
\begin{equation}
\gamma(\vec{x})=
\gamma_\parallel 
\left(1+\sum_\nu
\frac{\gamma_\perp^2}{(\omega_\nu-\omega_\mathrm{a})^2+\gamma_\perp^2}
\hspace{2pt}|a_{\nu0}|^2|\vec{E}_\nu|^2\right).
\label{eq:relax1}
\end{equation}
The second term in \eqref{relax1} is precisely the local enhancement of the atomic decay rate due to stimulated emission in the presence of the lasing fields. (A simplified \emph{spatially averaged} enhancement of the atomic decay rate  was previously  discussed in~\cite{vanExter1992}.) 

The Langevin force $f_\mu$ is the projection of the spontaneously emitted field onto the corresponding mode $\vec{E}_\mu$~\cite{Henry1986}. 
Defining $\vec{F}_\mu(t)\equiv\bold{F}_\mathrm{S}e^{i\omega_\mu t}$, the Langevin force $f_\mu$ is 
\begin{equation}
f_\mu(t)=\frac{i\displaystyle\int dx \vec{E}_\mu \cdot \vec{F}_\mu(t)  }
{\displaystyle\int dx(\omega_\mu^2\varepsilon)'_\mu\vec{E}_\mu^2}.
\label{eq:force}
\end{equation}

The full N-SALT TCMT equations \eqref{timeDelay} describe the most typical situation in laser dynamics of a ``class B" laser~\cite{Arecchi1984,Oppo1986,Lugiato1984}, in which the polarization of the gain medium can be adiabatically eliminated but the population dynamics is relatively slow and cannot be so eliminated.  However, much of the basic linewidth physics can be extracted from the limit when the population dynamics is also adiabatically eliminable, which describes ``class A" lasers. Since the mathematical analysis is simpler in this limit, we will begin the spectral analysis in \secref{spectrum}  with the latter  model.  We discuss this limit, which we refer to as the ``instantaneous model,"  in the following section.

\subsection{Instantaneous single-mode model}

When the population relaxation rate $\gamma(\vec{x})$ is (everywhere) large compared to the dynamical scales determining $a_\mu (t)$, the exponential terms in \eqref{timeDelay} act like  $\delta$ functions. After the spatial integration, and specializing in this section  to the single-mode case, we obtain the simple nonlinear  oscillator model driven by a weak Langevin force 
$f(t)$:
\begin{equation}
\dot{a}=C\left(a_0^2-|a|^2\right)a+f,
\label{eq:TCMT}
\end{equation}
where $C=\int dx \hspace{2pt}c(\vec{x})$ is the integrated nonlinear coupling. This instantaneous  nonlinear oscillator model was previously introduced by Lax~\cite{Lax1967a,Lax1966}, and has been used extensively in linewidth theories~\cite{Haken1984}. The N-SALT approach enables computing  the model's parameters \emph{ab initio}, taking full account of the spatial hole-burning term and the  vectorial nature of the fields [including  multimode effects, when generalizing \eqref{TCMT} to the multimode regime]. Also, our approach shows that  this well-known model can be explicitly
derived from the more general (non-instantaneous) model, presented in the previous section. Above the lasing threshold, $a_0 > 0$  and  $\mbox{Re}[C]>0$, and the system 
undergoes self-sustained oscillations with a stable steady state at $|a|=a_0$, as demonstrated in \figref{illustration}(b). 
In fact, near threshold one can show that $\mbox{Re}[C]$ is approximately the threshold gain, which balances
the cavity loss $\kappa$.  Hence the dynamical scale of $a(t)$ is of order $\kappa$, which must then be much
smaller than $\gamma(\vec{x})$ for the instantaneous model to hold; this is the standard dynamical condition for class A lasers~\cite{Arecchi1984,Oppo1986,Lugiato1984}.

The nonlinear term in \eqref{TCMT} and the multimode counterpart in   \eqref{coupling} are derived rigorously in appendix A, but we can motivate the resulting expressions using simple physical arguments. The nonlinear term  can be viewed as a shift in the oscillation frequency, i.e., $-i\Delta\omega=C(a_0^2-|a|^2)$. Using first-order perturbation theory~\cite{Raman2011},
the frequency shift due to a change in dielectric permittivity $\Delta\varepsilon$ is given by
\begin{equation}
\Delta\omega = 
- \omega_0^2 \frac{\displaystyle\int dx \hspace{2pt} \Delta \varepsilon \hspace{2pt} \vec{E}_0^2}
{\displaystyle\int dx (\omega_0^2\varepsilon)'_0\vec{E}_0^2}.
\end{equation}
Plugging in the differential of the permittivity due to small changes in the  squared mode amplitude,
$\Delta\varepsilon\approx \frac{\partial\varepsilon}{\partial |a|^2}(|a|^2-a_0^2)$, 
we find that the coupling coefficient in the instantaneous model is
\begin{equation}
C=\frac{-i\omega_0^2
\displaystyle\int dx \frac{\partial\varepsilon}{\partial |a|^2} \vec{E}_0^2}
{\displaystyle\int dx (\omega_0^2\varepsilon)'_0\vec{E}_0^2}.
\label{eq:InstC}
\end{equation}
This is the single-mode version of {\eqref{coupling} integrated over space due to rapid relaxation.  As we will see,
this simple result, combined with the spectrum of the Langevin noise (section IV), is all that is needed to derive the single-mode linewidth formula  \eqref{TCMTwidth}  (see Section V), and the multimode generalization also follows straightforwardly.  Hence, after analyzing the noise spectrum, we will first derive the linewidth within the instantaneous model before moving on to the more complicated case of the full N-SALT TCMT equations.  The latter will show that the basic linewidth formula is unchanged from that of the instantaneous model except for the addition of  side peaks due to the relaxation oscillations present in class B lasers.

\section{The autocorrelation function of the Langevin force \label{FDTsec}}

In this section, we express the  autocorrelation function of the Langevin force $f_\mu$
\begin{equation}
\left<f_\mu(t){f}_\nu^*(t')\right>=R_\mu\delta_{\mu\nu}\delta(t-t')
\label{eq:autoTime}
\end{equation}
 in terms of the autocorrelation function of the noise source  $\vec{F}_\mu$.   It is well known that quantum and thermal fluctuations can be modeled as zero-mean random variables, defined by their correlation functions~\cite{Lifshitz1980,Eckhardt1982}. This Rytov picture~\cite{Rytov1989} is essentially a consequence of the central-limit theorem (CLT)~\cite{Feller1945,Feller1968}, which holds since the classical forcing $\vec{F}_\mathrm{S}$ is the sum of a large number of randomly emitted photons. The autocorrelation function of $\vec{F}_\mathrm{S}$ can be found by invoking the fluctuation--dissipation theorem (FDT), as explained below.

 The probability distributions of the pumped medium and the electromagnetic   field obey Boltzmann statistics, with an effective local temperature $\beta$ defined in terms of the atomic inversion~\cite{Patra2000} (see definition in \secref{formula}). Under the typical conditions of local thermal equilibrium~\cite{Callen1951,Rytov1989,Dzyaloshinkii1961,Lifshitz1980,Eckhardt1982}, dissipation by optical absorption must be balanced by spontaneous emission from current fluctuations~$\vec{J}(\vec{x},t)$. One can then apply the  FDT for the Fourier-transformed forcing $\widehat{\vec{F}}_\mathrm{S}(\vec{x},\omega)=-4\pi\, i\omega\widehat{\vec{J}}(\vec{x},\omega)$~\cite{Henry1986Eq45}:
\begin{align}
\left<\widehat{\vec{F}}_\mathrm{S}(\vec{x},\omega)
\widehat{\vec{F}}\textsuperscript{*}\hspace{-5pt}_\mathrm{S}(\vec{x}',\omega')\right>=
\hspace{1.75in}\nonumber\\
2\hbar\omega^4\mbox{Im}\hspace{2pt}\varepsilon (\vec{x},\omega)
\coth\left(\frac{\hbar\omega\beta(\vec{x})}{2}\right)\delta(\vec{x}-\vec{x}')\delta(\omega-\omega').
\label{eq:FDT}
\end{align}

Using this result, we calculate the autocorrelation of the Langevin force $\widehat{f}_\mu$ [i.e., the Fourier transform of \eqref{force}, defined as~$\widehat{f}_\mu(\omega)\equiv \frac{1}{2\pi}\int_0^\infty dt\,e^{i\omega t}{f}_\mu(t)$] and we obtain
\begin{equation}
\left<\widehat{f}_\mu(\omega){\widehat{f}_\nu}^*(\omega')\right>=\widehat{R}_\mu(\omega)
\delta(\omega-\omega')\delta_{\mu\nu},
\label{eq:delta-auto}
\end{equation}
where the frequency-domain autocorrelation coefficient is  
\begin{equation}
\Scale[0.9]{\widehat{R}_\mu(\omega)=
4\hbar\omega^4\hspace{2pt}
\frac{\displaystyle\int\!\! dx\hspace{2pt} {|\vec{E}_\mu|}^2\mbox{Im}\hspace{2pt}\varepsilon(\omega)
\left[\frac{1}{2}\coth\left(\frac{\hbar\omega\beta}{2}\right)-\frac{1}{2}\right]}
{{\left|\displaystyle\int dx\hspace{2pt} \vec{E}_\mu^2 
(\omega_\mu^2\varepsilon)'_\mu\right|}^2}.
}
\label{eq:auto}
\end{equation}
 The $\frac{1}{2}$ factor subtracted from the hyperbolic cotangent is explained in appendix E and in~\cite{Henry1996}. 
 
The time-domain diffusion coefficient  $R_\mu$ can  be found directly from \eqref{auto} taking the inverse Fourier transform.  For the common case of a small linewidth, $\mbox{Im}\hspace{2pt}\varepsilon(\omega)$ and $\coth\left(\frac{\hbar\omega\beta}{2}\right)$ are nearly constant for  frequencies within the linewidth.  [This means, essentially, that the Langevin force $f_\mu(t)$ can be treated as white noise]. Consequently, one can approximate the diffusion coefficient in \eqref{auto} by its value at $\omega_\mu$. With this simplification, the time-domain diffusion coefficient in \eqref{autoTime} is  conveniently given by  $R_\mu = 2\pi \widehat{R}_\mu(\omega_\mu)$~\cite{Henry1986}.

More generally, however, including this frequency dependence corresponds to temporally correlated fluctuations, leading to non-Lorentzian corrections to the laser lineshape~\cite{Scully1988a,Scully1988b,Benkert1990a, Benkert1990b,Kolobov1993}. These ``memory effects'' can be addressed using our approach (as discussed in \secref{summary}) and we plan to  include them in future work.

\section{The laser spectrum \label{spectrum}}

In this section, we  calculate the laser spectrum using the N-SALT TCMT equations \eqsref{timeDelay}{TCMT} and the noise autocorrelation function \eqsref{delta-auto}{auto}. We begin by showing that  the phase of the lasing mode undergoes simple Brownian motion; consequently, the laser spectrum is a Lorentzian, with a width given by the phase-diffusion coefficient. In \secref{spectrum}.A, we calculate the phase-diffusion coefficient (hence the linewidth) for the instantaneous model \eqref{TCMT} and in \secref{spectrum}.B, we outline the analysis for the time-delayed model \eqref{timeDelay}, leaving the details of the derivation to  appendix B. More accurately, the spectrum of the time-delayed model consists of a central Lorentzian peak at the lasing resonance frequency and additional side peaks due to relaxation oscillations, which are present in class B lasers. The latter side peaks are the subject of \secref{spectrum}.C. 
\subsection{Instantaneous single-mode model \label{spectrum1}}
The complex mode amplitude $a(t)$ can be written in polar form as
\begin{equation}
a(t)=[a_0+\delta(t)]\hspace{2pt}e^{i\phi(t)}.
\label{eq:expandMode}
\end{equation}
$a_0$ is the steady-state  amplitude, while $\delta$ and $\phi$ are  \emph{real} amplitude and phase fluctuations. Substituting the modal expansion \eqref{expandMode} in \eqref{TCMT}, defining
\begin{align}
A\equiv2a_0^2\mbox{Re}\hspace{2pt}C\nonumber\\
B\equiv2a_0^2\mbox{Im}\hspace{2pt}C,
\label{eq:AB}
\end{align}
and keeping terms to first order in $\delta/a_0$, we obtain
\begin{align}
\dot{\delta} = -A\delta + f_\mathrm{R},
\label{eq:delta}\\
a_0\dot{\phi} = -B\delta + f_\mathrm{I},
\label{eq:phi}
\end{align}
where $f_\mathrm{R}\equiv\mbox{Re}\hspace{2pt}\{f\}$ and $f_\mathrm{I}\equiv\mbox{Im}\hspace{2pt}\{f\}$.
We  check the approximation of $|\delta|\ll a_0$ \emph{a posteriori} and we find that it generally holds (as was also shown in~\cite{Hui1993}), except near threshold ($a_0\rightarrow0$), which is a case we discuss in \secref{applications}.C.

When the nonlinear coupling coefficient is real ($B=0$), it is evident from \eqref{phi} that the phase undergoes simple Brownian motion (i.e., it is a Wiener process) and hence the phase variance increases linearly in time. An oscillator with Brownian phase noise has a Lorentzian spectrum~\cite{Demir2000}, and one can reproduce that result briefly as follows. The laser spectrum $S_\mathrm{a}(\omega)$ is given by the Fourier transform  of the autocorrelation function of $a(t)$:
\begin{equation}
\left<a(t)a^*(0)\right>\approx
a_0^2 
\left<e^{-i(\phi(t)-\phi(0))}  \right>
=
a_0^2 
e^{-\frac{1}{2}\left<\left(\phi(t)-\phi(0)\right)^2\right>} .
\label{eq:calc}
\end{equation}
For a Wiener phase, whose variance is 
$\left<(\phi(t)-\phi(0))^2\right>=\Gamma|t|$, the Fourier transform of \eqref{calc}  is a Lorentzian whose central-peak width is  $\Gamma$~\cite{Lax1967a}. 
In passing from the first to second step in \eqref{calc}, one neglects direct amplitude-fluctuation contributions  (which are decoupled from the phase) as these only introduce  broad-spectrum background noise, but do not affect the linewidth of the laser peak (we return to this point in \secref{spectrum}.C). In passing from the second  to the third step, one assumes that the phase is a Gaussian normal variable, which is justified as a consequence of the CLT.  

It is well known that also in the general case of $B\neq0$, the phase is a Wiener process, with a modified diffusion coefficient~\cite{Henry1982}. In order to calculate the phase variance explicitly, we solve  \eqsref{delta}{phi} and obtain
\begin{align}
\delta(t)=\int^t e^{-A(t-t')}f_\mathrm{R}(t')dt',
\label{eq:deltaSol}\hspace{0.5in}\\
a_0\phi(t)=-B \int^t \delta(t')dt' +
\int^t f_\mathrm{I}(t')dt'.
\label{eq:phiSol}
\end{align}
Substituting \eqref{deltaSol} into \eqref{phiSol}, using the autocorrelation function of $f$ \eqref{auto}, and performing the integration, one obtains that the  phase variance in the long-time limit is
$\left<(\phi(t)-\phi(0))^2\right> = 
\frac{R}{2a_0^2}\left(1+{\left(\frac{B}{A}\right)}^2\right)|t|$ (where terms growing more slowly than $|t|$ were neglected, as explained in greater detail in appendix B). Therefore, the linewidth is 
\begin{equation}
\Gamma=\frac{R}{2a_0^2}(1+\widetilde{\alpha}^2),
\label{eq:singleWidth}
\end{equation} 
where we have defined the generalized $\alpha$ factor: 
\begin{equation}
\widetilde{\alpha}=\frac{B}{A}=
\frac{\mbox{Im}\hspace{2pt}C}{\mbox{Re}\hspace{2pt}C},
\label{eq:alpha}
\end{equation}
with the nonlinear coefficient $C$ defined in \eqref{InstC}.
Substituting the autocorrelation function \eqref{auto} in \eqref{singleWidth} and using Poynting's theorem to relate $a_0^2$ to   the output power 
$P=\frac{\omega_0 a_0^2}{2\pi}\int_\mathrm{P} dx\hspace{2pt}(-\mbox{Im}\hspace{2pt}\varepsilon(\vec{x}))|\vec{E}_0(\vec{x})|^2$~\cite{Ge2010append}, we obtain the single-mode linewidth formula \eqref{TCMTwidth}. From \eqref{singleWidth}, it is evident that 
the Schawlow--Townes,  Petermann,  bad-cavity  and  incomplete-inversion factors are all included in the term $\frac{R}{2a_0^2}$, and generally  cannot be separated into the traditional factors of  \eqref{STP}~\cite{Pillay2014}.

When the nonlinear coupling coefficient is complex (i.e., when $B\neq0$), the resonance peak is not only broadened but is also shifted~\cite{Henry1982}. The shift in center frequency is found by keeping second-order terms in $\delta/a_0$ and calculating the average phase drift:
\begin{equation}
\delta\omega=\dot{\left<\phi\right>}=-\frac{RB}{4a_0^2A}.
\label{eq:shift}
\end{equation}
An identical formula was derived in~\cite{Henry1986} in a phenomenological instantaneous model. 

\figref{spectrumA} shows the spectrum of the instantaneous model, which is obtained by numerically  solving \eqref{TCMT} using a stochastic Euler scheme~\cite{Press2007}. Introducing the notation $\mathcal{F}(a)\equiv C(a_0^2-|a|^2)a$
and discretizing time as $a(n\Delta t) \approx a_n$, the Euler update equation for the $n$-th step is
\begin{equation}
a_n=a_{n-1}+\mathcal{F}\left(a_{n-1}\right)\Delta t+\sqrt{R\,\Delta t}\,\zeta,
\end{equation}
where $\zeta$ is a gaussian random variable of mean 0 and variance 1, i.e., $\zeta\in N(0,1)$. [For the data presented in \figref{spectrumA}, $\Delta t$ was decreased until the simulation results converged. In later sections (\figref{spectrumB}), we implemented a fourth-order Runge--Kutta method in order to achieve convergence]. The simulated spectra (noisy colorful curves) match the predicted Lorentzian lineshapes (solid black curves), which are calculated using \eqsref{singleWidth}{shift}. As $\widetilde{\alpha}$ increases, the linewidths are broadened and the center frequencies are shifted.

\begin{figure}
        \centering     
                 \includegraphics[width=0.325\textwidth]{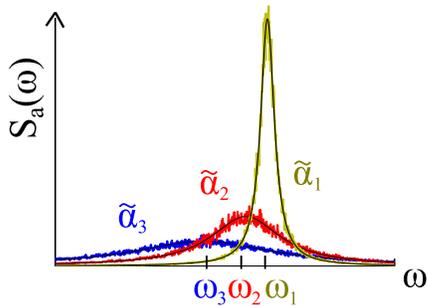}
	\caption{(Color online) Simulated spectrum $S_\mathrm{a}(\omega)$ of the instantaneous model \eqref{TCMT} with $\mbox{Re}\,C=10$, noise autocorrelation coefficient $R=0.1$ and three values of $\widetilde{\alpha}$: 10 (blue), 5 (red), 1 (yellow) ($C, R, S_\mathrm{a}$, and $\omega$ are given  in arbitrary frequency units). The noisy signal is the simulation result and the black curves are   Lorentzian lineshapes with widths $\Gamma$ and center frequency shifts    given by  \eqsref{singleWidth}{shift}. 
}
\label{fig:spectrumA}
  \end{figure}


\subsection{Time-delayed  multimode model \label{sperctrum2}}

We now turn to the laser spectrum produced by the time-delayed model, where the nonlinearity is dependent on the modal
amplitudes at previous times. Although we calculate the linewidth of the full time-delayed N-SALT TCMT equations \eqref{timeDelay} in appendix B, we begin this section by considering the  simplified case of a spatially homogeneous medium $\gamma(\vec{x}) \approx \gamma_0$ (this is  a good approximation for a uniformly pumped class B  laser operating near threshold). In this case, the single-mode time-delayed model  takes the form
\begin{equation}
\dot{a}=C\hspace{2pt}\left(\gamma_0\int^t dt' e^{-\gamma_0(t-t')}(a_0^2-|a(t')|^2)\right)a+f,
\label{eq:singleTimeDelay}
\end{equation}
where $C=\int dx \hspace{2pt}c(\vec{x})$ is the integrated nonlinear coupling  and $c(\vec{x})$ is defined in \eqref{coupling}.
This integro-differential equation can be turned into a first-order ODE by  using the modal expansion  from \secref{spectrum}.A: $a=(a_0+\delta)e^{i\phi}$, keeping terms to first order in $\delta/a_0$, and introducing the variable
\begin{equation}
\xi(t) = \gamma_0 \int^t dt' e^{-\gamma_0(t-t')}\delta(t').
\label{eq:xi}
\end{equation}
Then, \eqsref{singleTimeDelay}{xi} can be recast in the form $\dot{\vec{v}}=\mat{K}\vec{v}+\vec{f}$, where  $\vec{v}=\{\delta,a_0\phi,\xi\}$. 

However, most generally, the spatial dependence of $\gamma(\vec{x})$ cannot be neglected.  The time-averaged deviation $\xi(\vec{x},t)$ is therefore spatially dependent, and  one obtains an infinite-dimensional problem.  To simplify the algebra, we discretize space [e.g., discretizing \eqref{timeDelay} into a Riemann sum over sub-volumes $V_k$] and recover the continuum limit at the end. This yields the discrete-space multimode model:
\begin{align}
\dot{a}_\mu&=\nonumber\\
\sum_{\nu k}&C_{\mu\nu}^k \hspace{2pt}
\left(\gamma_k\int^t dt' 
e^{-\gamma_k(t-t')}(a_{\nu0}^2-|a_\nu(t')|^2)\right)a_\mu+f_\mu,
\label{eq:RiemannSum}
\end{align}
where the discretized nonlinear coupling coefficients are $C^k_{\mu\nu}=\int_{V_k} dx \hspace{2pt}c_{\mu\nu}(\vec{x})$ (so that $C_{\mu\nu}=\sum_k C^k_{\mu\nu}$), $\gamma_k$ is the relaxation rate at the $k$'th spatial point and $a_{\nu0}$ is the steady-state  amplitude of mode $\nu$.

In appendix B, we study the statistical properties of the solutions to \eqref{RiemannSum}. We introduce the  the M-dimensional vectors
 whose entries are $\Phi_\mu\!\equiv\!a_{\mu0}\phi_\mu$ (where $M$ is the number of active lasing modes)  and we calculate the covariance matrix $\left<{\Phi}_\mu(t){\Phi}_\nu(0)\right>$. We find that the result is independent of the relaxation rates $\gamma_k$ or the discretization scheme:
\begin{equation}
\left<\vec{\Phi}(t)\vec{\Phi}\hspace{2pt}^T(0)\right>=
\left(
\frac{\mat{R}}{2}+
\mat{B} \mat{A}^{-1}
\hspace{2pt} \frac{\mat{R}}{2} \hspace{2pt}
\left( \mat{B}\mat{A}^{-1}\right)^T\right)
\hspace{2pt} |t|.
\label{eq:multiAlpha}
\end{equation}
The matrices $\mat{A}$ and $\mat{B}$  correspond to the real and imaginary parts of the coupling matrices, with entries $A_{\mu\nu}=
2a_{\mu0}a_{\nu0}\mbox{Re}[C_{\mu\nu}]$
and $B_{\mu\nu}=
2a_{\mu0}a_{\nu0}\mbox{Im}[C_{\mu\nu}]
$. $\mat{R}$ is the autocorrelation function of the Langevin force vector $\vec{f}$ [defined in \eqref{auto}].  The diagonal of this matrix, divided by $|t|$ and by the squared modal amplitude, gives the generalized linewidths
\begin{equation}
\Gamma_{\mu}=
\frac{1}{2a_{\mu 0}^2}\left(
R_{\mu\mu}+
\left[\mat{B} \mat{A}^{-1}\mat{R}\left( \mat{B}\mat{A}^{-1}\right)^T\right]_{\mu\mu}\right).
\label{eq:multiWidth}
\end{equation}
Therefore, the generalized $\alpha$ factor (which is responsible for linewidth enhancement due to coupling of amplitude and phase fluctuations) is given by
\begin{equation}
\widetilde{\alpha}_\mu\equiv
\frac{1}{R_{\mu\mu}}
\left[\mat{B} \mat{A}^{-1}\mat{R}\left( \mat{B}\mat{A}^{-1}\right)^T\right]_{\mu\mu}.
\label{eq:generalAlpha}
\end{equation}
In the single-mode case ($M=1$), this matrix formula reduces to the single-mode linewidth of the instantaneous-model:
$\frac{R}{2a_0^2}(1+{\left(\frac{B}{A}\right)}^2)$ [\eqsref{singleWidth}{alpha} in \secref{spectrum}.A]. 

The linewidth in the time-delayed (class B) model is precisely the same (neglecting side peaks) as in the instantaneous (class A) model.  While this result was derived  for  single-mode class B semiconductor lasers using a  phenomenological rate-equation framework~\cite{vanExter1992}, we prove that this is generally the case in the multimode inhomogeneous regime. Naively, one might expect to obtain different linewidths due to the longer time over which  the fluctuations can grow. However, in appendix B we obtain a linewidth expression  which is independent of the relaxation-oscillation dynamics, which demonstrates that there is a cancellation of two competing processes: as $\gamma_\parallel$ decreases, amplitude fluctuations grow, but they are also averaged over  longer periods of time so that their effect is smaller.

\figref{spectrumB} presents the simulated spectrum of the time-delayed model in the homogeneous-$\gamma$ limit, which is obtained by numerically integrating \eqref{singleTimeDelay} (by applying a stochastic Euler scheme, as in \figref{spectrumA}). The width of the central peak of the spectrum matches our prediction \eqref{singleWidth}, independent of the value of $\gamma_0$. At intermediate relaxation rates, we also observe side peaks in the spectrum due to amplitude relaxation oscillations (RO), in addition to the central peak.

\begin{figure}
        \centering     
                 \includegraphics[width=0.5\textwidth]{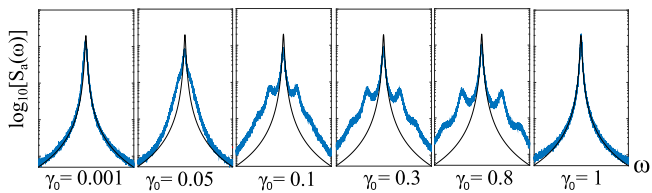}
	\caption{(Color online)  Simulated spectrum  of the time-delayed model  \eqref{singleTimeDelay} with $\mbox{Re}\,C=10$ and  $R=0.1$  (in arbitrary frequency units) at  six values of $\gamma_0$ (using a base 10 logarithmic scale for the $y$ axis).  The noisy signal is the simulation result and the black  curves are Lorentzian lineshapes with widths given by \eqref{singleWidth}.
}
\label{fig:spectrumB}
  \end{figure}

\subsection{Side peaks in the time-delayed model
\label{spectrum3}}

In class B lasers, amplitude fluctuations relax to steady state via relaxation oscillations~\cite{Siegman1986} and, consequently, give rise to side peaks in the spectrum, in analogy with amplitude modulation of harmonic signals. Mathematically, the oscillation arises from the second-order ODE generated by coupling of the $\dot{\delta}$ and $\dot{\xi}$ equations \eqsref{singleTimeDelay}{xi}, producing the coupled amplitude/gain oscillations.  Using the same methods that we applied to calculate the linewidth of the central resonance peak \eqref{multiWidth}, we also calculated the full side-peak spectrum in the multimode regime. Our formula is derived under the fairly general assumption that the central resonance peaks are narrower than the side peaks, which is the relevant regime for many lasers~\cite{vanExter1992}. Although the derivation uses the same techniques as in appendix B, it is fairly involved and will be provided in a subsequent manuscript~\cite{Pick2015}; we only summarize here.

As was shown in \secref{CMT}, far above threshold, the atomic relaxation rate \eqref{relax1} is enhanced  and can even be  dominated by the electromagnetic field. This modified relaxation rate, and in particular its spatial dependence due to hole-burning effects, has important implications on the RO spectrum which, to our knowledge, have not been treated before.   For simplicity, we focus here on the case of~$\alpha=0$. (Note that $\alpha$ factor effects on the RO spectrum have been observed and analyzed using a phenomenological \emph{homogeneous} time-delayed model in~\cite{vanExter1992}.)

In order to see how one can obtain a closed-form expression for the RO spectrum, recall that when calculating the spectrum of the central resonance peak in \secref{spectrum}.A, we neglected direct amplitude-fluctuation contributions in \eqref{calc}, i.e., in passing from the first to second step, we omitted a term of the form
\begin{equation}
\left<\delta(t)\delta(0)\right>\cdot
\left<e^{-i(\phi(t)-\phi(0))}  \right>.
\label{eq:omitted}
\end{equation}
Adding this term in \eqref{calc}, one finds that the full spectrum  consists of an additional term, which is given by the convolution of the real-amplitude fluctuation spectrum~$\left<\delta(t)\delta(0)\right>$   and the spectrum of the central resonance peak. In the instantaneous model, the amplitude autocorrelation function~$\left<\delta(t)\delta(0)\right>$ decays exponentially in time [see \eqref{deltaSol}] and the omitted term results in near-constant background noise. However, in the time-delayed model,
this neglected term is responsible for the RO side peaks.

For simplicity, consider first a model which can be solved straightforwardly; the single-mode homogeneous-$\gamma$  time-delayed model [i.e.,~$\gamma(\vec{x})\approx\gamma_0$ and~$\int dx \, c(\vec{x})=C$ as in \eqref{singleTimeDelay}], which describes 
uniformly pumped single-mode lasers near threshold. 
  Following the discussion in~\secref{spectrum}.B, we can rewrite~\eqsref{singleTimeDelay}{xi} as a set of linear equations and solve for~$\delta(t)$, obtaining
\begin{align}
\delta(t)=\int dt'  e^{-\frac{\gamma_0}{2}(t-t')}\hspace{2in}\nonumber\\
\hspace{0.1in}\times\left[
\cosh\left(\frac{\Delta}{2}(t-t')\right)+\frac{\gamma}{\Delta}\sinh\left(\frac{\Delta}{2}(t-t')\right)
\right]f_\mathrm{R}(t'),
\end{align}
where~$\Delta\equiv\sqrt{\gamma_0^2-4A\gamma_0}$. In the limit of well-resolved side peaks (e.g.,~$\frac{R}{a_0^2}\ll\gamma_0\ll A$), 
the amplitude autocorrelation function is approximately
\begin{equation}
\left<\delta(t)\delta(0)\right>\approx
\frac{R}{2}
\left[
\frac{\sin\sqrt{\gamma_0 A}\,t}{\sqrt{\gamma_0 A}}+\frac{\cos\sqrt{\gamma_0 A}\,t}{\gamma_0}
\right]
\times e^{-\frac{\gamma_0 t}{2}}.
\label{eq:sidebands}
\end{equation}
Thus, additional peaks in the spectrum arise at frequencies~$\omega_\mathrm{RO}=\omega_0\pm\sqrt{A\gamma_0}$  with widths~$\gamma_0$. In the high-Q limit near threshold, $A$ is proportional to the cavity decay rate $\kappa$, giving the expected behavior for the RO frequency. The side-peak amplitudes~$\frac{R}{4}\left[\frac{1}{\sqrt{\gamma_0 A}}+\frac{1}{\gamma_0}\right]$ diverge in the limit of $\gamma_0\rightarrow0$ (that is, when amplitude fluctuations are not small compared to the steady-state mode amplitude), but this is also the regime in which our analysis of the  spectrum  (\secref{spectrum}.A-B) breaks down. The inset in \figref{comparison}b shows  the simulated spectrum of the homogeneous time-delayed model \eqref{singleTimeDelay} (the same data was also shown in \figref{spectrumB}, but we include here the theoretical formula for the side-peak spectrum). The exact numerical solution of \eqref{singleTimeDelay} (blue curve) reproduces the analytic spectrum prediction of \eqref{sideBandFullSpec} (red curve). 

In the limits of extremely small/large relaxation rates $\gamma_0$ (compared to $A$), the side peaks disappear. In the former limit, they merge with the central resonance peak and in the latter case, they merge with the background noise.  This behavior can be explained by inspection of the $\dot{\delta}$ and $\dot{\xi}$ equations \eqsref{singleTimeDelay}{xi} in the appropriate limits.  When the relaxation rate is very large, the time-delayed model reduces to the instantaneous model, which represents the case where the atomic population follows the field adiabatically. In the opposite limit of extremely small relaxation, the field follows the atomic population adiabatically. In other words, a clear separation of atomic and optical time scales will result in the absence of RO side peaks.

\begin{figure}
\centering
                 \includegraphics[width=0.5\textwidth]{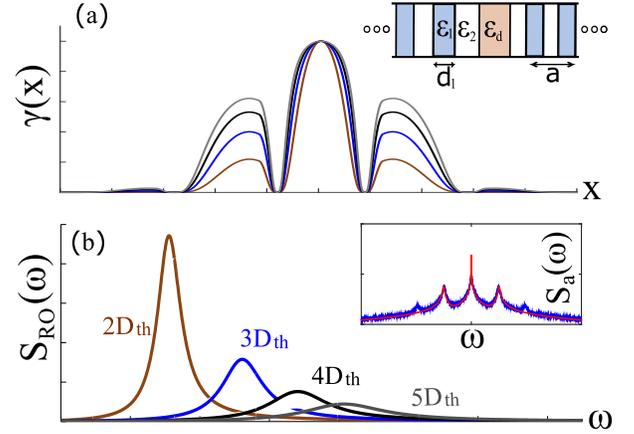}
		\caption[]{(Color online) Dressed decay rate and the RO spectrum based on SALT solutions  of a 1d PhC laser. 
								Inset: a quarter-wave  PhC (period $a=1\mbox{ mm}$ and  alternating layers with permittivities $\varepsilon_1=16+0.1i$ and $\varepsilon_2=2+0.1i$ and thicknesses $d_1=\frac{a \sqrt{\varepsilon_2}}{\sqrt{\varepsilon_1}+\sqrt{\varepsilon_2}}$ and $ d_2=a-d_1$). The  center region has permittivity $\varepsilon_\mathrm{d}=3+0.1i$ and contains gain atoms with bandwidth $\gamma_\perp=3\mbox{ mm}^{-1}$ and resonance frequency $\omega_\mathrm{a}=25\mbox{ mm}^{-1}$.
(a) Dressed decay $\gamma(\vec{x})$  evaluated using \eqref{relax1} at five pump values ($2D_\mathrm{th}$ brown, $3D_\mathrm{th}$ blue, $4D_\mathrm{th}$ black, and $5D_\mathrm{th}$ gray). 
		(b) Side-peak spectrum $S_\mathrm{RO}(\omega)$ evaluated using \eqref{sideBandFullSpec} for the five pump values of (a). 
		Inset: full simulated spectrum $S_\mathrm{a}(\omega)$ on a semi-log scale (of base 10) of the homogeneous  time-delayed model \eqref{singleTimeDelay} with $\gamma_0=0.09, A=10, B=0, R=0.01$ (in arbitrary frequency units). The noisy signal is the simulation result and the red curve is the theoretical lineshape \eqref{sideBandFullSpec}.}
\label{fig:comparison}
\end{figure}

In the most general  spatially inhomogeneous time-delayed model, the full spectrum takes the simple form
\begin{align}
&S_\mathrm{a}(\omega)=
\Scale[1.3]{\frac{\Gamma}{\omega^2+\left(\frac{\Gamma}{2}\right)^2}}+\nonumber\\
&\Scale[0.85]{\displaystyle
\frac{\Gamma}
{\omega^2\left[1-\!\displaystyle\int \!\!dx \frac{A(\vec{x})\gamma(\vec{x})}{\omega^2+(\frac{\Gamma}{2}+\gamma(\vec{x}))^2}
\right]^2+
\left[\displaystyle\int \!\!dx \frac{A(\vec{x})\gamma(\vec{x})\left(\gamma(\vec{x})+\frac{\Gamma}{2}\right)}{\omega^2+\left(\gamma(\vec{x})+\frac{\Gamma}{2}\right)^2}\right]^2}
\label{eq:sideBandFullSpec}}
\end{align}
where $A(\vec{x})$ is the real part of the local nonlinear coupling [defined in \eqref{AB}], $\gamma(\vec{x})$ is the effective  decay rate, and $\Gamma$ is the central peak linewidth.   (This formula is valid when the central resonance peak is narrower than the side peaks $\Gamma\ll\gamma_\parallel$.) Like our linewidth formula, this formula is easy to evaluate via spatial integrals of the SALT solutions. 

While the homogeneous time-delayed model near threshold agrees with standard results on relaxation oscillations~\cite{vanExter1992}, the full model above threshold, combined with SALT, is able to include effects not contained in other treatments.  As the pump is increased far above threshold, the effects of stimulated emission strongly increase the atomic relaxation rate, and spatial hole burning causes that rate $\gamma(\vec{x})$ to vary substantially in space [see \eqref{relax1}].   These two effects cause both a shift and a broadening of the side peaks compared to the near-threshold result.  \figref{comparison}  shows the dressed decay rate $\gamma(\vec{x})$ and the side-peak spectrum $S_\mathrm{RO}(\omega)$ [as given by the second term of  \eqref{sideBandFullSpec}], based on a SALT calculation of a one-dimensional photonic crystal (PhC) laser, at four different pump values well above threshold. 
[The pump value is controlled via the parameter $D_\mathrm{p}$ in \eqref{d-dot}, and we denote the threshold value of $D_\mathrm{p}$ by $D_\mathrm{th}$.]
 This type of cavity (depicted in the inset of \figref{comparison}a) supports a single mode at the simulated parameter regime, which is localized near the defect region. (Further discussion of this structure is given in \secref{applications}.A below.) As can be seen from \figref{comparison}a, the decay rate $\gamma(\vec{x})$ is enhanced at high intensity regions (i.e., near the defect), and it increases further  as the pump increases. \figref{comparison}b demonstrates the shifting and broadening of the side peaks. 
\section{The generalized $\alpha$ factor \label{applications}}

\begin{figure*}[t]
        \centering     
                 \includegraphics[width=0.8\textwidth]{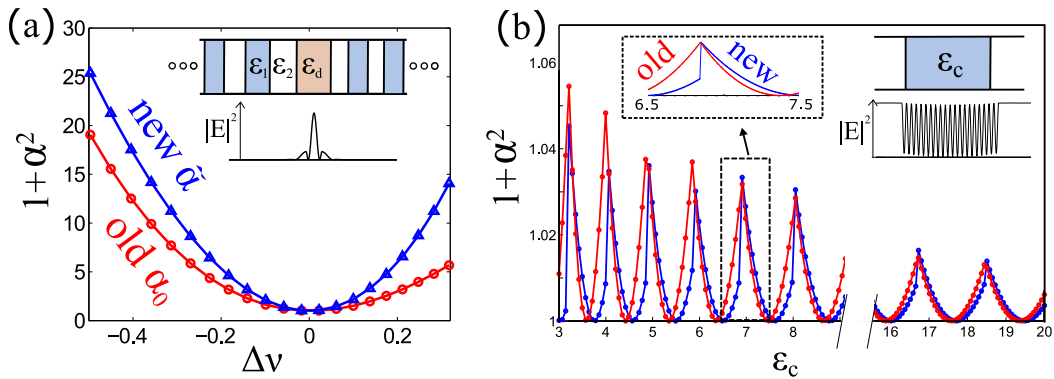}
                  \caption{(Color online) 
(a) The generalized (blue) and traditional (red) $\alpha$ factors of a PhC laser vs.  relative detuning  $\Delta\nu\equiv\frac{\omega_0-\omega_\mathrm{a}}{\omega_0}$.
Upper inset: quarter-wave  PhC geometry (see  caption of  \figref{comparison}). The gain parameters are  
$\gamma_\perp=3\mbox{ mm}^{-1}$ and a varying  $\omega_\mathrm{a}$. 
Lower inset: intensity distribution of the lasing mode. 
(b) $\alpha$ factor for an open cavity laser vs. passive permittivity $\varepsilon_\mathrm{c}$. 
Blue (red): generalized (traditional) $\alpha$ factor.
Upper inset: 
dielectric slab, of permittivity $\varepsilon_\mathrm{c}$, bounded by air on both sides, containing gain atoms with $\omega_\mathrm{a}=15\mbox{ mm}^{-1}$, $\gamma_\perp=3\mbox{ mm}^{-1}$.
Lower inset:
intensity distribution of the lasing mode. 
Leftmost inset: enlarged segment of the main plot, around $\varepsilon_\mathrm{c}=7$.  
}
\label{fig:SingleHenry}
  \end{figure*}

Our TCMT derivation of the linewidth formula yields a generalized $\alpha$ factor \eqref{generalAlpha} which depends on the eigenmodes  $\vec{E}_\mu(\vec{x})$ and eigenfrequencies $\omega_\mu$ of the full nonlinear SALT equations. This is an advance over previous linewidth formulas; the  \emph{ab-initio} scattering-matrix  linewidth formulas did not obtain an $\alpha$ factor~\cite{Chong2012,Pillay2014}, whereas other traditional laser theories that derived $\alpha$ factors could not  handle the full nonlinear equations~\cite{Kuppens1994B}. Therefore,  in the following section, we focus on the generalized $\alpha$ factor. We compare the generalized and traditional factors in \secref{applications}.A, and then we evaluate these factors in the single-mode (\secref{applications}.B) and multimode (\secref{applications}.C) regimes.

\subsection{Comparison with traditional $\alpha$ factor
\label{applications1}}

Linewidth broadening  due to amplitude--phase coupling (that is, the $\alpha$ factor linewidth enhancement) was first studied in the 1960s by Lax in the context of  single-mode detuned gas  lasers~\cite{Lax1966}. The Lax $\alpha$ factor is $1+\alpha_0^2$, where $\alpha_0$ is the normalized detuning of the lasing frequency from the atomic resonance, i.e., $\alpha_0=\frac{\omega_0-\omega_\mathrm{a}}{\gamma_\perp}$, which is equal to the ratio of the real part of the gain permittivity to its imaginary part, or equivalently the ratio $\frac{\mbox{Re}\,\Delta n_\mathrm{g}}{\mbox{Im}\,\Delta n_\mathrm{g}}$, where $\Delta n_\mathrm{g}$ is the refractive index change due to fluctuations in the gain. 
Two decades later, Henry derived an amplitude--phase coupling enhancement factor of the same general type in semiconductor lasers~\cite{Henry1982}, $\alpha_0 = \frac{\mbox{Re}\,\Delta n_\mathrm{g}}{\mbox{Im}\,\Delta n_\mathrm{g}}$, but in the latter case these refractive-index changes arise from carrier-density fluctuations and take a different form. Here, we are considering atomic gain media, so our $\alpha$ factor generalizes the Lax form.
 
The difference between our single-mode generalized $\alpha$ factor \eqref{alpha} and that of Lax arises because
we take into account spatial variation in the gain permittivity due to spatial hole-burning and also the non-Hermitian (complex) nature of the lasing mode.  Hence we expect our factor to reduce to the Lax factor in some limits.
For instance, consider the situation that was discussed in the last paragraph of \secref{formula}  of a low-loss 1d Fabry-P{\'e}rot cavity laser, operating near threshold. In this case, the nonlinear coupling coefficient  is approximately 
$C\approx \frac{-i\omega_0}{2\varepsilon}\frac{\int \frac{\partial\varepsilon }{\partial|a|^2}\vec{E}_0^2}{\int \vec{E}_0^2}$,
 and one can show that the generalized $\alpha$ factor is $\widetilde{\alpha}=\frac{\mbox{Im}\hspace{2pt}C}{\mbox{Re}\hspace{2pt}C}\approx\frac{\mbox{Re}\hspace{2pt}\Delta\varepsilon}{\mbox{Im}\hspace{2pt}\Delta\varepsilon}$ (the last approximation is valid since in essentially all realistic cavities, the modes can be chosen to be predominantly real, i.e., have small imaginary parts).

In many cases, however, our $\widetilde{\alpha}$  deviates from the traditional factor $\alpha_0$.  An obvious  example is when the lasing frequency precisely coincides with the atomic resonance frequency. In this case, the traditional factor vanishes, but $\widetilde{\alpha}$ does not necessarily vanish. In the next section, we calculate and discuss the characteristic properties of the generalized $\alpha$ factor for two 1d laser structures.

\subsection{Generalized single-mode $\alpha$ factor \label{applications2}}

In this section, we evaluate the differences between the generalized and traditional $\alpha$ factors in 1d model systems. We solve the full nonlinear SALT equations using our recent finite-difference frequency-domain (FDFD) SALT solver~\cite{Esterhazy2013}.

The generalized factor $\widetilde{\alpha}$ can deviate significantly from the traditional factor $\alpha_0$ when the latter is large (a similar argument was made in~\cite{Duan1990}). To see this, let us   write the nonlinear coupling coefficient qualitatively as $C\propto(1+i\alpha_0)(1+i\beta)$, where the  term  $1+i\alpha_0$ is  associated with the atomic lineshape $\frac{\gamma_\perp}{\omega_0-\omega_\mathrm{a}+i\gamma_\perp}$,
 and the  term  $1+i\beta$ is a  complex factor due to the remaining integral factors 
(we refer to the latter term as the \emph{modal contribution} to the $\alpha$ factor). Typically $\beta\ll1$ and, consequently, the generalized factor is approximately  $\widetilde{\alpha}\approx\alpha_0+\beta(1+\alpha_0^2)$, so the difference between the generalized and traditional factors grows quadratically with $\alpha_0$.   
  
To verify this argument, we study a model system in which the magnitude of $\alpha_0$ can be controlled. Consider a quarter-wave dielectric photonic crystal (PhC), with a defect at the center of the structure (the geometry is depicted in the upper inset of \figref{SingleHenry}a, similar to the structure that was studied in \figref{comparison}). Adding enough layers of  the periodic structure on each side of the defect to mimic an infinite structure, one finds that the system has a localized mode in the vicinity of the defect (lower inset), whose resonance frequency is fixed to a real value within the energy gap~\cite{Jannopoulos2008}. To study finite-threshold lasers, we introduce gain and some passive loss (i.e., a positive imaginary permittivity term, which pushes the resonance poles away from the real axis in the complex plane). Since the resonance frequency of the defect mode is fixed by the geometry, by varying the resonance frequency of the gain, we control the detuning of the lasing mode from the atomic resonance, thus controlling the size of $\alpha_0$. As demonstrated in the figure,  the deviation $|\widetilde{\alpha}-\alpha_0|$ grows as the detuning $\Delta\nu\equiv\frac{\omega_0-\omega_\mathrm{a}}{\omega_0}$ increases.

The openness of the cavity also results in an enhancement of the $\alpha$ factor; the more open it is, the larger is the necessary imaginary part of the lasing mode, which causes a deviation from the standard formula.
In order to test this prediction, we evaluate the generalized $\alpha$ factor for an open-cavity laser (\figref{SingleHenry}b), where we can control the radiative loss rate through the cavity walls and, consequently, this part of the modal contribution to 
$\widetilde{\alpha}$.  We consider a  cavity  which consists of a dielectric slab (with permittivity $\varepsilon_\mathrm{c}$)  surrounded by air  on both sides, with gain spread homogeneously inside the slab (upper rightmost inset). The reflectivity of the cavity walls is determined by the difference  in cavity and air permittivities $\Delta\varepsilon=\varepsilon_\mathrm{c}-\varepsilon_0$. For relatively small dielectric mismatch, the cavity is relatively low-$Q$ and our $\alpha$ factor differs significantly from the Lax factor.
As $\Delta\varepsilon$ increases and the cavity~$Q$ increases, the generalized $\alpha$ factor converges to the original factor, so that the red and blue curves in the figure overlap. 

\begin{figure*}[t]
        \centering     
                 \includegraphics[width=0.8\textwidth]{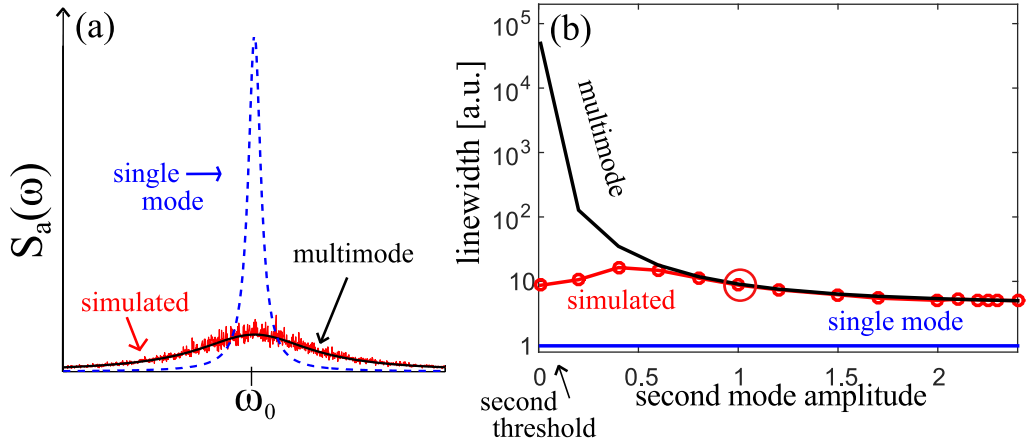}
	\caption{(Color online) (a) Spectrum  near a resonance peak in the presence of an   additional mode. 
	Numerical simulations of \eqref{TCMT} (red curve) and analytic single-mode (blue) and multimode (black) formulas. The simulation parameters are chosen so that there are two lasing modes with the same steady-state amplitudes $a_{k0} = 1$ and diffusion coefficients $R_{kk} = 0.05$, and with substantial cross correlations:
$C_{kk} = 5, C_{kl} = 4+4i, k\ne l$ (in arbitrary frequency units). (b) Linewidth of central resonance peak  vs. output power in the neighboring mode [$a_{10}=1$ and $a_{20}\in(0,3)$]. Simulated spectrum (red) and analytic single-mode and multimode  formulas  (blue and black curves). The point $a_{10}=a_{20}=1$ is encircled, and corresponds to the parameter values of \figref{MultiHenry}a.}
\label{fig:MultiHenry}
\end{figure*}

Unlike a photonic-crystal defect-mode cavity where there is a finite bandwidth of confinement~\cite{Jannopoulos2008}, this dielectric cavity has an infinite number of possible lasing resonances and thus when we sweep $\Delta\varepsilon$, the $\alpha$ factor peaks periodically. This is because the free spectral range of the cavity  is $\Delta\omega\approx\frac{2\pi}{\sqrt{\varepsilon_\mathrm{c}}L}$~\cite{Saleh2007} and, therefore, changing $\varepsilon_\mathrm{c}$  corresponds to shifting the passive resonances and, consequently, the lasing modes. Every time a lasing mode crosses an atomic  resonance, $\alpha_0$ vanishes  and correspondingly $\widetilde{\alpha}$ becomes very small.  The traditional factor is maximized when the atomic resonance is equidistant from two passive modes. The peak value is proportional to the free spectral range and, therefore, we find that it is proportional to $1/\sqrt{\varepsilon_\mathrm{c}}$. 
This type of effect may not have been observed previously because in macroscopic cavities, the cavity resonances are very dense on the scale of the gain bandwidth, so the lasing mode can never be substantially detuned.  However, in microcavities with large free spectral range, this could be an important effect.
Another intriguing property of the generalized $\alpha$ factor is that it varies discontinuously at the peaks (as is shown more clearly in the upper left-most inset). The traditional factor $\alpha_0$ depends only on the mode detuning from resonance, so it approaches the same value on different sides of the peak. In contrast, the generalized factor $\widetilde{\alpha}$ depends on the mode profile $\vec{E}_\mu$, which differs between the two interchanging laser modes on different sides of the peak, producing the observed asymmetry.

\subsection{Generalized multimode $\alpha$ factor \label{applications3}}
Our {\it multimode} linewidth formula includes linewidth corrections from neighboring modes, which enter through the generalized $\alpha$ factor (since phase fluctuations in each of the modes couple to amplitude fluctuations in all other  modes due to saturation of the gain). According to the traditional ST formula \eqref{STP}, when phase  cross-correlations between different modes are neglected, each resonance-peak width is inversely proportional to the corresponding modal output power. We find that  when phase  cross-correlations are included, the linewidth of each mode is a sum of inverse output powers of all the other modes---a type of multimode Schawlow--Townes relation. To see how this comes about, recall that the generalized $\alpha$ factor, as given by \eqref{generalAlpha}, is proportional to  $\left[\mat{B} \mat{A}^{-1}\hspace{2pt} \mat{R}\hspace{2pt}\left( \mat{B}\mat{A}^{-1}\right)^T\right]_{ii}$. We show in appendix C that individual factors in the product scale as  $[\mat{B}\mat{A}^{-1}]_{ij}\propto\frac{a_{i0}}{a_{j0}}$, where $a_{j0}$ is the steady-state amplitude of the $j$'th mode. Therefore, the multimode $\alpha$ factor  is proportional to the sum:~$
a_{i0}^2\sum_j{\frac{\left(\mbox{const}\right)\times R_{jj}}{a_{j0}^2}}$, i.e., a sum over terms which scale as inverse   output powers.

In the two-mode case,
the linewidth formula for a lasing mode in the presence of a neighboring mode is given explicitly by
\begin{align}
&\Gamma_{1}=\frac{R_{11}}{2a_{10}^2}+\nonumber\\
&\frac{R_{11}}{2a_{10}^2}
\left[\frac{C_{11}^\mathrm{I}C_{22}^\mathrm{R}-C_{21}^\mathrm{I}C_{21}^\mathrm{R}}
{C_{11}^\mathrm{R}C_{22}^\mathrm{R}-C_{12}^\mathrm{R}C_{21}^\mathrm{R}}\right]^2\!+\!
\frac{R_{22}}{2a_{20}^2}
\left[
\frac{C_{11}^\mathrm{R}C_{12}^\mathrm{I}-C_{11}^\mathrm{I}C_{12}^\mathrm{R}}
{C_{11}^\mathrm{R}C_{22}^\mathrm{R}-C_{12}^\mathrm{R}C_{21}^\mathrm{R}}\right]^2,
\label{eq:2mode}
\end{align}
where $C^\mathrm{R}_{ij}\equiv\mbox{Re}\hspace{2pt}C_{ij}$ and $C^\mathrm{I}_{ij}\equiv\mbox{Im}\hspace{2pt}C_{ij}$. 
(A similar expression was derived  in~\cite{Elsasser1985}, by using a phenomenological version of the two-mode TCMT equations.)
As predicted by the multimode ST relation,  
the last term in \eqref{2mode} is inversely proportional  to the output power of the second mode $a_{20}^2$. This term  becomes significant  when the power in the first mode greatly exceeds the power in the second mode (i.e., when $P_1\gg P_2$), correcting the unrealistic Schawlow--Townes prediction that the linewidth vanishes when $P_1\rightarrow\infty$; a similar argument was made in~\cite{Elsasser1985}. \figref{MultiHenry}a presents the spectrum of a two-mode instantaneous model \eqref{TCMT} in the parameter regime where   cross-correlations between the two modes are significant. 
   The linewidth of the simulated spectrum (red curve) is in complete agreement with the generalized formula \eqref{2mode} (black curve), but deviates substantially from the single-mode formula \eqref{singleWidth} (blue curve). In order to reach the regime where this deviation is substantial, in practice, one needs to design a cavity in which the two lasing modes have comparable amplitudes and detunings from the atomic resonance frequency.  

Eq. \eqref{2mode} predicts an unphysical divergence near the second threshold, i.e., when $a_{20}\rightarrow0$ (see black curve in \figref{MultiHenry}b). In retrospect, this singularity is to be expected, since the assumptions of our derivation break down  in this limit. (Note that an equivalent divergence was present in~\cite{Elsasser1985}.) In calculating the phase variance, we assumed that amplitude fluctuations in all modes were small compared to the steady-state amplitudes ($\delta_\mathrm{I}\ll a_{i0}$), and this assumption is no longer valid near threshold. The N-SALT TCMT equations \eqref{timeDelay} are still valid, however---it is only their analytical solution for $\left<\vec{\Phi}\vec{\Phi}^T\right>$ that is problematic. Therefore, we study the threshold regime numerically, via stochastic simulations of the N-SALT TCMT equations. As shown in \figref{MultiHenry}b, the simulated linewidth of the first mode approaches a finite value near the second threshold (red curve), and  this value is significantly larger than the linewidth prediction one obtains when neglecting the second mode (blue curve). Even at the  threshold, noise in the second mode mixes with the first mode through  off-diagonal nonlinear coupling terms, thus increasing the linewidth. 

Linewidth enhancement at the thresholds of neighboring lasing-modes suggests that the linewidth must also be enhanced \emph{below} the modal thresholds [in the regime where radiation from non-lasing modes is incoherent, commonly called amplified spontaneous emission (ASE)]. We believe that this phenomenon could be explored using a future generalization of our formalism, with some modifications (extending earlier work~\cite{Hui1993,Lax1967a} on linewidth enhancement from ASE).

\section{Full-vector 3d example\label{3d}}

In order to illustrate the full generality of our approach, we apply it in this section to study a three-dimensional photonic-crystal (PhC) laser.
The steady-state properties of this system (i.e., the lasing threshold and mode characteristics) were previously explored in~\cite{Esterhazy2013}.  We use those solutions here to calculate the laser linewidth  [using \eqref{TCMTwidth}], and we compare the relative contributions of the various correction  factors. 

The simulated PhC consists of a dielectric slab patterned by  a hexagonal lattice  of air holes (\figref{3d}a). A defect is introduced by decreasing the radii of seven holes at the center of the structure~\cite{Lin2001}, giving rise to a doubly-degenerate mode which is situated at the defect (spatially) and in the bandgap of the lattice (spectrally). We select the TE-like mode out of the degenerate  pair by imposing even and odd reflection symmetry at $x = 0$ and $y = 0$ respectively, as well as an even reflection symmetry at $z = 0$. Staying close to a potential experimental realization, we choose the pump profile to be uniform inside the high-index dielectric near the defect region, and zero elsewhere. We solve the SALT equations using our scalable FDFD solver, and track the evolution of the first lasing mode upon increasing the pump strength from zero to five times the first-threshold value.

\begin{figure}
\centering
                 \includegraphics[width=0.5\textwidth]{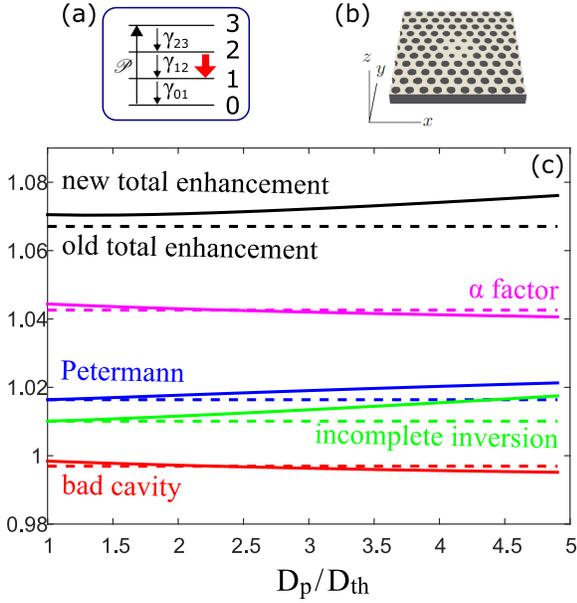}
		\caption[]{(Color online) Linewidth correction factors for a 3d PhC laser. 
		  (a) The PhC consists of a hexagonal lattice of air holes (with period $a = 1\mbox{ mm}$ and radius $0.3\mbox{ mm}$) in a dielectric medium with index $n =\sqrt{\varepsilon_\mathrm{c}}=3.4$.  The slab has a thickness of $0.5\mbox{ mm}$, with air above and below, terminated by PML absorbers. A cavity is formed by seven holes of radius of $0.2 \mbox{ mm}$. The pump is non-zero in the hexagonal region for height $2 \mbox{mm}$ in the z-direction. (Borrowed from \cite{Esterhazy2013}.) 
		  (b) Schematics of a 4-level gain medium. Levels $\left|1\right>$ and $\left|2\right>$ form the lasing transition, with resonance frequency $\omega_\mathrm{a} = 1.5 \mbox{ mm}^{-1}$ and polarization decay 
$\gamma_\perp= 2.0 \mbox{ mm}^{-1}$. The population decay rates are $\gamma_{01}/\gamma_{12}=\gamma_{23}/\gamma_{12}=10^{2}$ and $\gamma_{12}/\gamma_\perp=10^{-2}$. The pump rate $\mathscr{P}$ is varied in the range $\mathscr{P}/{\gamma_{12}}=0.4\hdots2.2$.
(c) Traditional (dashed) and generalized (solid) correction  factors, as defined in \tabref{Factors}. The total correction is defined as the product of the (traditional and generalized) Petermann, $\alpha$, bad-cavity, and incomplete-inversion factors. The $x$ axis is the $D_\mathrm{p}/D_\mathrm{th}$, where $D_\mathrm{p}$ is the SALT effective pump parameter (see text) and $D_\mathrm{th}$ is the effective threshold pump.}
\label{fig:3d}
\end{figure}

Typically, realistic laser structures do not use 2-level gain media, but employ a more complex optical scheme which involves multiple levels  and transitions in order to achieve significant inversion and depletion of the ground-state population. In this section, we apply our formalism to a 4-level gain medium (\figref{3d}b), using a generalization of SALT~\cite{Cerjan2012}, which finds the stationary multimode lasing properties of an $N$-level gain  medium.  As shown in~\cite{Cerjan2012},  an $N$-level system can be mapped into an effective 2-level system, which obeys the (2-level) SALT equations with renormalized pump ($D_\mathrm{p}$) and atomic relaxation rates ($\gamma_\parallel$). Consequently, the linewidth of a 4-level laser will be given by our generalized formula \eqref{TCMTwidth} with the appropriately renormalized coefficients. By choosing the decay rate between the lasing transition levels ($\gamma_{12}$ in \figref{3d}b) to be much smaller than the decay rates into the upper ($\gamma_{23}$) and out of the lower ($\gamma_{01}$) states, we can achieve substantial inversion and ground-state depletion. Consequently, the incomplete-inversion factor is approximately  $n_\mathrm{sp}\approx1$, close to typical measured values~\cite{Kuppens1994}.



\figref{3d}c presents the traditional and new correction factors (dashed and solid lines respectively), as defined in \tabref{Factors}. 
We find that those factors are relatively small for this system and, consequently, the  deviations between the new and traditional factors are small. A small Petermann factor arises since the first lasing mode has a relatively high quality factor (i.e., the cold-cavity resonance pole is at $\omega_0=1.725-0.00512i\mbox{ mm}^{-1}$ with a quality factor of $Q\approx700$, in agreement with experimental realization~\cite{Lin2001}). Moreover, the cold-cavity resonance lies well within the gain bandwidth, resulting in small $\alpha$ and bad-cavity corrections. The generalized  factor $\widetilde{\alpha}$ (solid purple line) is obtained from from \eqsref{InstC}{alpha}. Deviations of $\widetilde{\alpha}$  from the traditional factor $\alpha_0\equiv\frac{|\omega_0-\omega_\mathrm{a}|}{\gamma_\perp}$ (dashed purple line) are due to modal contributions to the $\alpha$ factor (see \secref{applications}.B). The generalized Petermann factor (full  blue curve) is compared against the traditional factor (dashed blue line), which is expressed in terms of  the SALT mode (instead of the passive cavity mode). The cavity region is taken to be the entire high-index medium. (Note the the generalized and traditional factors agree at threshold). Both the Petermann and $\alpha$ factors increase the linewidth. However, the generalized  and traditional  bad-cavity  factors (full and dashed red curves respectively)  lead to linewidth reduction.   

Last, we evaluate the incomplete-inversion factor  $\widetilde{n}_\mathrm{sp}$. The inversion $D(x)$ is found from the SALT solutions of the effective 2-level system. The excited state population $N_{2}(x)$ can be derived straightforwardly, using the results of~\cite{Cerjan2012} as follows. Assuming that the populations in the non-lasing levels $\left|0\right>$ and $\left|3\right>$ are at  steady-state, one can express those populations in terms of the populations in the lasing transition $\left|1\right>$ and $\left|2\right>$. Then, by invoking  the density conservation condition, $\sum_{i}N_\mathrm{I}=n$, where $n$ is the atom number density and $N_\mathrm{I}$ are the individual level populations, one finds that the population in $\left|2\right>$ is given by
\begin{align}
N_2=\frac{n+\tau D}{1+\tau},
\label{eq:POP2}
\end{align}
where $\tau\equiv 1+\frac{2\gamma_{01}}{\gamma_{23}}+\frac{\gamma_{01}}{\mathscr{P}}$. 
Having obtained expressions for $D$ and for $N_2$, we have all that is needed to calculate the incomplete-inversion factor $\widetilde{n}_\mathrm{sp}$. We define the ``linear incomplete-inversion factor'' ($n_\mathrm{sp}$ dashed green line) as the ratio $\frac{N_2(D_\mathrm{p})}{D_\mathrm{p}}$, [i.e., both the excited-state population \eqref{POP2} and the inversion are evaluated at $D=D_\mathrm{p}$, neglecting hole-burning effects]. The ``nonlinear incomplete-inversion factor'' ($\widetilde{n}_\mathrm{sp}$ solid green line) is defined in \tabref{Factors}.  
The nonlinear factor $\widetilde{n}_\mathrm{sp}$ coincides with the linear factor $n_\mathrm{sp}$ at threshold, but  exceeds the traditional factor at higher pumps. We also plot the total linewidth correction, which is defined as the product of the (traditional and new) Petermann, $\alpha$, bad-cavity, and incomplete-inversion factors.

\section{Concluding Remarks \label{summary}}

We presented a generalized multimode linewidth formula, obtained from the N-SALT TCMT equations for the lasing mode amplitudes, which we derived starting  from the Maxwell--Bloch equations and using the fluctuation--dissipation theorem to determine the statistical properties of the  noise. Our generalized linewidth formula \eqref{TCMTwidth} reduces to the traditional formula \eqref{STP} for low-loss cavities and simple lasing structures, but deviates significantly from the traditional theories for high-loss wavelength-scale laser cavities.  By basing our derivation on the SALT steady-state lasing modes, it is possible to apply our formula to cavities of arbitrarily complex geometry (e.g., photonic crystal or microdisk lasers~\cite{He2013,Painter1999,Loncar1999,Park2004}) and arbitrary openness (e.g. random lasers~\cite{Tureci2008}).  Also, since SALT includes to high accuracy the effects of spatial hole-burning, our formula includes both gain saturation and the spatial variation of the gain permittivity well above threshold, plus all effects due to modal couplings. From a computational point of view it is important to point out that our formula is analytical and can be evaluated immediately from the output of a numerical SALT calculation without any significant computational effort. 
A manuscript describing a brute-force numerical validation of our theory against numerical solution of the Maxwell--Bloch equations is currently being prepared~\cite{Cerjan2014}.
Given only the laser geometry, the pumping profile, and characteristic properties of the gain (i.e., its resonance frequency $\omega_\mathrm{a}$ and decay rate $\gamma_\perp$), our formula enables linewidth calculation, including a generalized $\alpha$-factor and accounting for temperature variations, 
at a level of generality that was not possible before. This generality is most important, of course, in cases where the new result is substantially different than previous theories, and it would be interesting to study laser cavities  in which the discrepancy is as large as possible.

One such case is that of lasers which contain exceptional points (EPs) in their spectrum, which are points of degeneracy where 
two (or more) eigenfrequencies and eigenfunctions coalesce~\cite{Heiss2010,Moiseyev2011}. EPs in laser systems have been explored recently, both theoretically~\cite{Liertzer2012} and experimentally~\cite{Brandstetter2014}. At the EP, the modes become self-orthogonal and that  causes the denominator of \eqref{TCMTwidth} to vanish and is already known to greatly enhance the Petermann factor~\cite{Lee2008}. Since a similar denominator appears in the integrals defining our generalized $\alpha$ factor \eqsref{coupling}{alpha}, we expect that our $\widetilde{\alpha}$ will differ substantially from previous results near an EP (and similarly for the inhomogeneous-temperature correction).

An important and exciting addition to the theory would be a treatment of amplified spontaneous emission (ASE) from modes below threshold; we believe this can be achieved by deriving TCMT equations for below-threshold (passive) modes, in which there is no steady-state oscillation (generalizing previous ASE work which used simplified  models~\cite{Lax1967a,Hui1993}). 
Incorporating the ASE contribution to the spectrum will allow us to follow the noise through the lasing thresholds, correcting the unphysical divergence which was discussed in \secref{applications}.B. More importantly, treating below threshold ASE should allow an ab-initio theory of LEDs in arbitrary cavities

Future work could also incorporate several additional corrections that were not treated in this paper.  Our derivation applies to isotropic materials described by a scalar permittivity $\varepsilon$, but extension to anisotropic permittivity $\hat{\varepsilon}$, magnetic permeability ($\hat{\mu}$), and even bianisotropic materials would be very straightforward (e.g., for an anisotropic $\hat{\varepsilon}$, the only change is that $\varepsilon \vec{E}^2$ factors and similar are replaced by $\vec{E} \cdot (\hat{\varepsilon} \vec{E})$ etcetera, as in~\cite{Arnaud1986}). As discussed in \secref{spectrum}.C, we are also able to exploit our framework to analytically solve for the relaxation-oscillation side-peak spectra, and are currently preparing a manuscript presenting this analysis~\cite{Pick2015}. We believe it will be possible to extend our formalism to handle non-Lorentzian lineshapes arising from frequency dependence (correlations) in the noise within the laser linewidth~\cite{Scully1988a,Scully1988b,Benkert1990a, Benkert1990b,Kolobov1993}, as also discussed in \secref{FDTsec}. Instead of treating the noise spectrum $S_\vec{F}(\omega)$ as a constant $S_\vec{F}(\omega_\mu)$, one needs to include a first-order correction , e.g., by Taylor expanding $S_\vec{F}(\omega)$ around $\omega_\mu$; it might be convenient to fit $S_\vec{F}(\omega)$ to a Lorentzian matching the amplitude and slope at $\omega_\mu$, since the Fourier transform of a Lorentzian is an exponential that should be easy to integrate.
Finally, as noted above, although our derivation was for the two-level Maxwell--Bloch equations, a similar approach should apply to more complex gain media (including multi-level atoms~\cite{Cerjan2012}, multiple lasing transitions, and gain diffusion~\cite{Cerjan2014b}.) The N-SALT linewidth theory can be generalized to account for these laser models following along the lines of our approach here. 

\begin{acknowledgments}
This work was partially supported by the Army Research Office through the Institute for Soldier Nanotechnologies under Contract No. W911NF-13-D-0001. ADS and AC acknowledge the support of NSF Grant No. DMR-1307632. CYD acknowledges the support  of Singapore NRF Grant No.~NRFF2012-02. The authors would like to thank Bo Zhen, Aristeidis Karalis, Amir Rix,  Owen Miller, and Homer Reid for helpful discussions. 

\end{acknowledgments}


\appendix
\appendixsec{Derivation of  N-SALT TCMT  }{appendixA}

In this appendix, we derive the TCMT equations for the lasing mode amplitudes. Our starting point is the Maxwell--Bloch equations~\cite{Haken1984,Lamb1964}, which describe the dynamics of the electromagnetic field in a resonator interacting with a two-level gain medium:
\begin{equation}
\nabla\times\nabla\times \vec{E}+\varepsilon_\mathrm{c}\hspace{2pt}\ddot{\vec{E}}
=-4\pi\ddot{\vec{P}}+\vec{F}_\mathrm{S},
\label{eq:e-dot}
\end{equation}
\begin{equation}
\dot{\vec{P}}=-i(\omega_\mathrm{a}-i\gamma_\perp)\vec{P}-\frac{i\gamma_\perp}{4\pi} \vec{E} D,
\label{eq:p-dot}
\end{equation}
\begin{equation}
\dot{D}=-\gamma_\parallel
\left[D_\mathrm{p}-D+2\pi i
(\vec{E}\cdot\vec{P}^*-\vec{E}^*\cdot\vec{P})\right],
\label{eq:d-dot}
\end{equation}
where $\vec{E}$ is the electromagnetic field, while $\vec{P}$ and $D$ are the atomic polarization and population inversion. (From here on, for brevity, we refer to $D$ as the ``inversion.'') $\omega_\mathrm{a}$ is the atomic resonance frequency, and $\gamma_\perp$ and $\gamma_\parallel$ are the population and inversion relaxation rates. $D_\mathrm{p}$ is the external pump, which determines the steady-state inversion, and $\varepsilon_\mathrm{c}$ is the passive dielectric permittivity.  The field, polarization and inversion are measured in their natural units: $e_c=p_c=\hbar\sqrt{\gamma_\parallel\gamma_\perp}/(2g)$ and $d_c=\hbar\gamma_\perp/(4\pi g^2)$ respectively, where $g$ is the atomic dipole matrix element~\cite{Tureci2006,Ge2010,Esterhazy2013}. We introduce spontaneous emission noise by including a random source term $\vec{F}_\mathrm{S}=4\pi\frac{\partial\vec{J}}{\partial t}$ in \eqref{e-dot}, written in  the frequency domain as
\begin{equation}
\widehat{\vec{F}}_\mathrm{S}(\vec{x},\omega)=-i4\pi\omega\widehat{\vec{J}}(\vec{x},\omega),
\label{eq:SE noise}
\end{equation}
where $\widehat{\vec{J}}(\vec{x},\omega)$ is a random fluctuating current, and the correlations of  $\widehat{\vec{F}}_\mathrm{S}(\vec{x},\omega)$ are given by the FDT.

Steady-state ab-initio laser theory (SALT)  handles the noise-free regime of the Maxwell--Bloch equations (i.e., $\widehat{\vec{F}}_\mathrm{S}=0$) and reduces this set of coupled equations to a  frequency-domain nonlinear generalized eigenvalue problem for the electric field $\widehat{\vec{E}}$ (as reviewed in Sec. 1.1).  When noise is introduced ($\widehat{\vec{F}}_\mathrm{S}\neq0$), the cavity field is  perturbed from steady-state and the nonlinear permittivity is modified (Sec. 1.2). This gives rise to a restoring force (denoted $\widehat{\vec{F}}_\mathrm{NL}$), which we calculate in Sec. 1.3. The noise-driven field $\widehat{\vec{E}}$ is then found by integrating the  Green's function (derived in Sec. 1.4) over the noise terms $\widehat{\vec{F}}_{S}$ and $\widehat{\vec{F}}_\mathrm{NL}$.   Finally,  the TCMT equations are obtained by transforming back into the time domain (Sec. 1.5).

\subsection{Review of SALT}

We begin by reviewing the steady-state theory.  In the SALT approach, the steady-state electromagnetic field is expressed as a superposition of a finite number of lasing modes:
\begin{equation}
\vec{E}_0(\vec{x},t) = \sum_\mu  \vec{E}_\mu(\vec{x}) a_{\mu0} e^{-i\omega_\mu t},
\end{equation}
where $\vec{E}_0(\vec{x},t)$ denotes the steady-state field and  $a_{\mu0}$ are the  steady-state modal amplitudes. 
The lasing modes $\vec{E}_\mu(\vec{x})$ are real frequency solutions of the nonlinear eigenvalue problem
\begin{equation}
\left[\nabla\times\nabla\times -\omega_\mu^2\widehat{\varepsilon}_0(\omega_\mu,a_0)\right] \vec{E}_\mu(\vec{x})=0,
\label{eq:steady ME}
\end{equation}
with outgoing boundary conditions. The  effective permittivity  has a linear (passive) term  $\varepsilon_\mathrm{c}$ and a nonlinear ($\vec{E}$-dependent) gain term:
\begin{equation}
\widehat{\varepsilon}_0(\omega,a_0)= \varepsilon_\mathrm{c}+
\frac{\gamma_\perp}{\omega-\omega_\mathrm{a}+i\gamma_\perp}
D_0(a_0).
\label{eq:epsi SALT}
\end{equation}
The steady-state inversion $D_0(a_0)$ [which is a notation shortcut for $D_0(\{\vec{E}_\mu\},\{\omega_\mu\},\{a_{\mu0}\})$] is  given by
\begin{equation}
D_0(a_0)=
\frac{D_\mathrm{p}}{1+
\sum_\mu 
\frac{\gamma_\perp^2}{(\omega_\mu-\omega_\mathrm{a})^2+\gamma_\perp^2}|a_{\mu0}|^2|\vec{E}_\mu|^2}.
\label{eq:D steady}
\end{equation}
To avoid possible confusion, note that in previous SALT works, the steady-state inversion was denoted by $D$ and $D_0$ was the external pump parameter, whereas in this work, $D_0$ is the steady-state inversion and $D_\mathrm{p}$ is the external pump parameter.

\subsection{Noise-driven Maxwell-Bloch equations}

In the presence of a  small noise source, the electric field and polarization can be written as superpositions of the  steady-state lasing modes with time-dependent amplitudes $a_\mu(t)$ and $b_\mu(t)$:
\begin{align}
\vec{E}(\vec{x},t)= \sum_\mu \vec{E}_\mu(\vec{x})a_\mu(t)e^{-i\omega_\mu t}\nonumber\\
\vec{P}(\vec{x},t)= \sum_\mu \vec{P}_\mu(\vec{x})b_\mu(t)e^{-i\omega_\mu t}.
\label{eq:ansatz}
\end{align}
 Substituting the perturbation  ansatz \eqref{ansatz} into the polarization equation \eqref{p-dot}, we obtain
\begin{equation}
(\dot{b}_\mu+i\omega_\mu b_\mu)\vec{P}_\mu=
-i(\omega_\mathrm{a}-i\gamma_\perp)b_\mu \vec{P}_\mu-\frac{i\gamma_\perp a_\mu }{4\pi}\vec{E}_\mu D.
\label{eq:p-step}
\end{equation}
Taking the Fourier transform and rearranging terms, we find
\begin{equation}
\widetilde{B}_\mu \vec{P}_\mu
=\frac{1}{4\pi} \frac{\gamma_\perp}{\omega-\omega_\mathrm{a}+i\gamma_\perp}\widehat{a}_\mu*\widehat{D}\vec{E}_\mu,
\label{eq:pol}
\end{equation}
where we have introduced the shifted frequency $\omega\equiv\omega_\mu+\Omega$
 and the Fourier-domain envelopes $\widehat{a}_\mu(\Omega)=\widehat{a}_\mu(\omega-\omega_\mu)$, $\widetilde{B}_\mu(\Omega)$ and $\widehat{D}(\Omega)$. The asterisk * denotes a convolution.

 Next, consider Eq. \eqref{e-dot} in the frequency domain 
 \begin{equation}
\nabla\times\nabla\times \widehat{\vec{E}}-\omega^2\varepsilon_\mathrm{c}\hspace{2pt}(\widehat{\vec{E}}+4\pi\widehat{\vec{P}})
=\widehat{\vec{F}}_\mathrm{S}.
\label{eq:e-dot-FD}
\end{equation}
When the spacing between adjacent lasing modes is much larger than their  linewidths, a noise source with frequency $\omega\approx\omega_\mu$ excites only the mode $\vec{E}_\mu(\vec{x})$. Equivalently, the Green's function can be approximated by the contribution of the single pole at $\omega_\mu$. 
(Note that we require only that the peaks in the laser spectrum above threshold are non-overlapping; we do not require  isolated resonances in the passive cavity spectrum.) Therefore, at frequencies $\omega\approx\omega_\mu$, we can substitute  \eqref{pol} into \eqref{e-dot-FD} and obtain an effective equation for the noise-driven field $\widehat{\vec{E}}_\mu(\vec{x},\omega)$:
\begin{equation}
\left[\nabla\times\nabla\times-\omega^2
\varepsilon(\omega,a)\right]\widehat{\vec{E}}_\mu(\vec{x},\omega)
=\widehat{\vec{F}}_\mathrm{S}(\vec{x},\omega),
\label{eq:ME-modeMu}
\end{equation}
where the effective permittivity $\varepsilon(\omega,a)$ is given by
\begin{equation}
\varepsilon(\omega,a) \widehat{\vec{E}}_\mu(\vec{x},\omega)=
\left[\varepsilon_\mathrm{c} \widehat{a}_\mu+\frac{\gamma_\perp}{\omega-\omega_\mathrm{a}+i\gamma_\perp}\widehat{D}*\widehat{a}_\mu\right]
\vec{E}_\mu(\vec{x}).
\label{eq:epsi}
\end{equation}
The second variable of $\varepsilon(\omega,a)$ denotes the implicit  dependence of $\varepsilon$ on the modal amplitudes  $a_\mu$  through the Fourier transform of the inversion $\widehat{D}$. 
We calculate $\widehat{D}$ explicitly in the next section.
  
\subsection{The atomic inversion}

The noise source $\widehat{\vec{F}}_\mathrm{S}$ perturbs the modal amplitudes $a_\nu$  from steady state, causing a change in the atomic inversion $D$. We neglect dispersion corrections  to $D$ (which amounts to setting  $\dot{b}_\mu=0$ in \eqref{p-step}~\cite{Lugiato1984}) as these corrections  do not affect the linewidth formula to leading order in the noise [see discussion following \eqref{epsiBody} in the main text]. From \eqref{d-dot} and \eqref{p-step}, we obtain
\begin{align}
\Scale[1.05]{
\dot{D}=-\gamma_\parallel 
\left[D-D_\mathrm{p}+
\sum_\nu
\frac{\gamma_\perp^2}{(\omega_\nu-\omega_\mathrm{a})^2+\gamma_\perp^2}
\hspace{2pt}
|a_\nu|^2D\hspace{2pt}|\vec{E}_\nu|^2\right]}
\label{eq:D}.
\end{align}
In order to solve \eqref{D}, we linearize the time dependent products $|a_\nu|^2D$ in the sum  around the steady state $|a_\nu|^2D\approx a_{\nu0}^2D_{0}+
D_{0}(|a_\nu|^2-a_{\nu0}^2)+a_0^2(D-D_{0})$, where $D_{0}$ is the steady state (SALT) inversion \eqref{D steady}.
To simplify the notation, we define the local decay rate
\begin{equation}
\gamma(\vec{x})\equiv
\gamma_\parallel 
\left(1+\sum_\nu
\frac{\gamma_\perp^2}{(\omega_\nu-\omega_\mathrm{a})^2+\gamma_\perp^2}
\hspace{2pt}|a_{\nu0}|^2|\vec{E}_\nu|^2\right).
\label{eq:relax}
\end{equation}
The second term in \eqref{relax} gives precisely the increased atomic decay rate due to stimulated emission.
Using the definitions above , \eqref{D} becomes
\begin{align}
\dot{D}
&=-\gamma(\vec{x})(D-D_0)\nonumber\\
&-\gamma_\parallel D_0 
\cdot
\sum_\nu 
\frac{\gamma_\perp^2}{(\omega_\nu-\omega_\mathrm{a})^2+\gamma_\perp^2}
\hspace{2pt}
|\vec{E}_\nu|^2(|a_\nu|^2-|a_{\nu0}|^2),
\label{eq:stepD}
\end{align}
which we can integrate, and obtain
\begin{align}
D =  D_0+\sum_\nu 
D_{0} \left(\frac{\gamma_\perp^2}{(\omega_\nu-\omega_\mathrm{a})^2+\gamma_\perp^2}
|\vec{E}_\nu|^2\right)\times\nonumber\\
\gamma_\parallel
\int^t dt' e^{-\gamma(\vec{x})(t-t')}
(|a_{\nu0}|^2-|a_\nu(t')|^2).
\label{eq:D-res}
\end{align}

Having derived an explicit expression for $D(t)$, we substitute its Fourier transform $\widehat{D}$ into the effective permittivity \eqref{epsi} and obtain
\begin{equation}
\varepsilon(\omega,a)\widehat{\vec{E}}_\mu\approx
\varepsilon(\omega,a_0)\widehat{\vec{E}}_\mu+\sum_\nu \chi_\nu(\omega,a_0)
\widehat{\Delta a_\nu}*\widehat{\vec{E}}_\mu,
\label{eq:permExpand}
\end{equation}
where
$\varepsilon(\omega,a_0)$ is the steady-state SALT permittivity which was defined in  \eqref{epsi SALT},
$\chi_\nu(\omega,a_0)$ is the permittivity differential due to deviation in the modal amplitude $a_\nu$
[which we denote by ``$\frac{\partial\varepsilon}{\partial|a|^2}$'' in the text, e.g., in \eqref{coupling}]:
\begin{align}
\chi_{\nu}\equiv \frac{\gamma_\perp}{\omega-\omega_\mathrm{a}+i\gamma_\perp}
 D_{0} 
\left(\frac{\gamma_\perp^2}{(\omega_\nu-\omega_\mathrm{a})^2+\gamma_\perp^2}
|\vec{E}_\nu|^2\right)
\frac{\gamma_\parallel}{\gamma(\vec{x})},
\label{eq:Deps}
\end{align}
and $\widehat{\Delta a}_\nu$ is the Fourier transform of the time-averaged modal deviation from steady state
\begin{align}
\Delta a_\nu=
\gamma(\vec{x})
\int^t dt' e^{-\gamma(\vec{x})(t-t')}
(|a_{\nu0}|^2-|a_\nu(t')|^2).
\label{eq:Da}
\end{align}

Substituting the permittivity expansion \eqref{permExpand} into Maxwell's equation \eqref{ME-modeMu}, we obtain
\begin{equation}
\Scale[0.95]{
\left[\nabla\times\nabla\times-\omega^2
\varepsilon(\omega,a_0)\right]\widehat{\vec{E}}_\mu(\vec{x},\omega)
=\widehat{\vec{F}}_\mathrm{NL}(\vec{x},\omega)
+\widehat{\vec{F}}_\mathrm{S}(\vec{x},\omega)},
\label{eq:ME-eff}
\end{equation}
where the nonlinear restoring force is 
\begin{equation}
\widehat{\vec{F}}_\mathrm{NL}(\vec{x},\omega)
=\omega^2
\sum_\nu \chi_\nu(\omega,a_0)\widehat{\Delta a_\nu}*\widehat{\vec{E}}_\mu(\vec{x},\omega).
\label{eq:FNL}
\end{equation}
The left-hand side of \eqref{ME-eff} is just the  linearized steady-state equation \eqref{steady ME}, and the  nonlinear correction to the effective permittivity due to the noise $\widehat{\vec{F}}_\mathrm{S}$ appears as an additional  source term $\widehat{\vec{F}}_\mathrm{NL}$.
As noted above, the noise-driven field $\widehat{\vec{E}}_\mu$ is found by integrating the Green's function of the steady-state equation \eqref{steady ME} over the noise terms $\widehat{\vec{F}}_\mathrm{NL}$ and $\widehat{\vec{F}}_{S}$. In the following section we derive an approximate formula for the Green's function.

\subsection{The linearized steady state Green's function }

The single-pole approximation of the Green's function is valid for frequencies near the resonances $\omega\approx\omega_\mu$ as long as the spectrum consists of non-overlapping resonance peaks, i.e.,  when the spacing between resonant modes exceeds the modal linewidths. First, let us rewrite the left-hand side of   \eqref{steady ME} as an operator $\mathcal{L}_\omega$ acting on the field $\vec{E}(\vec{x},\omega)$:
 \begin{equation}
\mathcal{L}_\omega \vec{E}(\vec{x},\omega)\equiv
\left(\nabla\times\nabla\times -\omega^2\widehat{\varepsilon}_0(\omega,a_0)\right)
\vec{E}(\vec{x},\omega).
 \end{equation}
Next, we choose a complete set (see below) of eigenfunctions  $\vec{E}_n(\vec{x},\omega)$ and eigenvalues $\lambda_n(\omega)$   of the operator $\mathcal{L}_\omega$:
\begin{equation}
\mathcal{L}_\omega  \vec{E}_n(\vec{x},\omega)=\lambda_n(\omega)  \vec{E}_n(\vec{x},\omega).
\label{eq:eigenvalue problem}
\end{equation}
We define the inner product of two vector fields, $\vec{A}(\vec{x})$  and $\vec{B}(\vec{x})$,  as $(A,B)\equiv\int dx \hspace{3pt}\vec{A}(\vec{x})\cdot \vec{B}(\vec{x})$. The operator $\mathcal{L}_\omega$ is complex symmetric under this inner product, i.e.,  $(A,\mathcal{L}_\omega B)=(\mathcal{L}_\omega  A, B)$~\cite{Jannopoulos2008,Moiseyev2011}. Therefore, we use unconjugated inner products throughout the derivation. In order to treat the set~$\{\vec{E}_n\}$ as a discrete (countable) basis, a convenient theoretical trick is to place the system in a box with absorbing boundary layers in which the absorption turns on more and more gradually. This procedure also gives the states $\vec{E}_n$ finite norms $(E_n,E_n)$. Because the operator is non-Hermitian, completeness of the basis can break down at an ``exceptional point''~\cite{Heiss2010,Moiseyev2011}, but exceptional points are not generically present—they must be forced by careful tuning of parameters. Therefore, we assume completeness in this manuscript and will treat the influence of exceptional points (self-orthogonal modes) as a limiting case in a future paper, as discussed in \secref{summary}.

Let $G(\omega,\vec{x},\vec{x}')$ be the Green's function of the operator $\mathcal{L}_\omega$, defined via  $\mathcal{L}_\omega G(\omega,\vec{x},\vec{x}')= \delta(\vec{x}-\vec{x}')$~\cite{Arfken2006B}.  Given the complete set of eigenfunctions and eigenvalues $\{\vec{E}_n,\lambda_n\}$, the Green's function can be expressed as the sum~\cite{Arfken2006B}
\begin{equation}
G(\omega,\vec{x},\vec{x}')=\sum_n \frac{ \vec{E}_{n}(\vec{x}) \vec{E}^T_{n}(\vec{x}')}
{\lambda_n(\omega)\cdot\displaystyle\int dx \hspace{3pt}\vec{E}^2_n(\vec{x}) }.
\label{eq:Green all}
\end{equation}
Each lasing mode is associated with an eigenvalue $\lambda_\mu(\omega)$ of $\mathcal{L}_\omega$, which has a zero at a real frequency $\omega= {\omega}_\mu$. Consequently, $G(\omega,\vec{x},\vec{x}')$ has a pole at $ {\omega}_\mu$ and at frequencies near $\omega_\mu$, it is dominated by a single term in the sum. Expanding $\lambda_\mu(\omega)$ around the pole $\lambda_\mu(\omega)\approx (\omega- {\omega}_\mu)\lambda_\mu' $ (where $\lambda_\mu'\equiv\left.\frac{\partial \lambda}{\partial\omega}\right|_{\omega_\mu}$), we obtain
\begin{equation}
G_{\mu\mu}(\omega,\vec{x},\vec{x}') \approx 
\frac{ \vec{E} _{\mu}(\vec{x}) \vec{E} _{\mu}^T(\vec{x}')}
{ (\omega- {\omega}_\mu)  \lambda_\mu'\cdot\displaystyle\int dx \hspace{3pt}\vec{E}_\mu^2(\vec{x})   }.
\label{eq:G before sub}
\end{equation}
In order to evaluate $\lambda_\mu'$, let us  rewrite  $\mathcal{L}_\omega$ as $\mathcal{L}_\omega\approx \mathcal{L}_{ {\omega}_\mu}+V(\omega)$, where 
$\mathcal{L}_{ {\omega}_\mu}\equiv\nabla \times \nabla \times - \omega_\mu^2 \widehat{\varepsilon}_0( {\omega}_\mu)$ and 
$V(\omega)\equiv -   \left[\omega^2\widehat{\varepsilon}_0(\omega)\right]_\mu'   (\omega- {\omega}_\mu)$. 
According to  the Hellmann--Feynman theorem, the derivative of the eigenvalue $\lambda_\mu(\omega)$ with respect to $\omega$ is given by 
\begin{equation}
\lambda_\mu'  = 
\frac{\displaystyle\int dx \hspace{3pt} \vec{E} _\mu^2(\vec{x}) 
\left[-\omega^2\widehat{\varepsilon}_0(\omega)\right]_\mu'}
{\displaystyle\int dx \hspace{3pt}\vec{E}_\mu^2(\vec{x})},
\label{eq:HF}
\end{equation}
and substituting   \eqref{HF} in   \eqref{G before sub}, we find that for frequencies near the resonances $\omega\approx\omega_\mu$, the Green's function is approximately
\begin{equation}
G_{\mu\mu}(\vec{x},\vec{x}',\omega)\approx \frac{\vec{E}_\mu(\vec{x})\vec{E}^T_\mu(\vec{x}')}
{(\omega_\mu-\omega)\displaystyle\int dx \hspace{2pt}\vec{E}_\mu^2(\vec{x})
 \left[\omega^2\widehat{\varepsilon}_0(\omega)\right]_\mu'}.
 \label{eq:single pole}
\end{equation}

\subsection{The N-SALT TCMT equations}
Having derived an expression for the Green's function, the noise-driven field can be found by integrating the Green's function over the source terms $\widehat{\vec{F}}_\mathrm{NL}(\vec{x}',\omega)$ and $\widehat{\vec{F}}_\mathrm{S}(\vec{x}',\omega)$:
\begin{align}
\widehat{\vec{E}}_\mu(\vec{x},\omega)
&=\sum_\nu{\omega}_\mu^2\int dx' G(\vec{x},\vec{x}',\omega)\chi_\nu(\omega_\mu,a_0)\widehat{\Delta a_\nu}*\widehat{\vec{E}}_\mu\nonumber\\
 &+
\int dx' G(\vec{x},\vec{x}',\omega) \widehat{\vec{F}}_\mathrm{S}.
\label{eq:beforeSUB}
\end{align}
In the first term on the right-hand side, we approximate $\omega\approx\omega_\mu$ because the correction term is $\mathcal{O}\left[(\omega-\omega_\mu)\cdot\widehat{\Delta a}\right]$, which is second order in the noise. 
Substituting the single-pole approximation \eqref{single pole} in \eqref{beforeSUB} yields
\begin{align}
\vec{E}_\mu\widehat{a}_\mu&=\sum_\nu
\frac{\omega_\mu^2}{(\omega-\omega_\mu)}
\frac{\vec{E}_\mu\int \chi_\nu(\omega_\mu,a_0)\vec{E}_\mu^2}{\displaystyle\int dx \hspace{2pt}\vec{E}_\mu^2(\vec{x})
 \left[\omega^2\widehat{\varepsilon}_0(\omega)\right]_\mu'}\widehat{\Delta a_\nu}*\widehat{a}_\mu
 \nonumber\\
 &+
\frac{\vec{E}_\mu}{\omega-\omega_\mu}
\frac{\int \widehat{\vec{F}}_\mathrm{S}(\vec{x}',\omega)\vec{E}_\mu}{\displaystyle\int dx \hspace{2pt}\vec{E}_\mu^2(\vec{x})
 \left[\omega^2\widehat{\varepsilon}_0(\omega)\right]_\mu'}.
\end{align}
Finally, multiplying both side by $\omega-\omega_\mu$ and taking the inverse Fourier transform, we arrive at the N-SALT TCMT equations, which govern the evolution of the modal amplitudes $a_\mu$: 
\begin{align}
&\dot{a}_\mu=
\sum_\nu
\int dx
c_{\mu\nu}(\vec{x})
\nonumber\\
&\gamma(x)
\int^t dt' e^{-\gamma(\vec{x})(t-t')}
(|a_{\nu0}|^2-|a_\nu(t')|^2)
    a_\mu+
f_\mu(t).
\end{align}
The nonlinear coupling coefficient is
\begin{equation}
c_{\mu\nu}(\vec{x})\equiv
-i\omega_\mu^2 
\frac{\chi_\nu(\omega_\mu,a_0) \vec{E}_\mu^2 
}
{\displaystyle\int dx \hspace{2pt}\vec{E}_\mu^2(\vec{x})
 \left[\omega^2\widehat{\varepsilon}_0(\omega)\right]_\mu'},
\label{eq:Cderiv}
\end{equation}
and the Langevin force is
\begin{equation}
f_\mu(t)\equiv
i\frac{\int [{\vec{F}}_\mathrm{S}(\vec{x}',t)e^{-i\omega_\mu t}]\vec{E}_\mu}
{\displaystyle\int dx \hspace{2pt}\vec{E}_\mu^2(\vec{x})
 \left[\omega^2\widehat{\varepsilon}_0(\omega)\right]_\mu'}.
\end{equation}

\appendixsec{Linewidth of the multimode time-delayed model}{appendixB}

In this section, we calculate the laser linewidth for the multimode time-delayed model by generalizing the solution strategy of \secref{spectrum} in the text. We begin our analysis with the discretized time-delayed N-SALT TCMT equation \eqref{RiemannSum} (repeated here for convenience):
\begin{align}
\dot{a}_\mu= \hspace{3in}\nonumber\\
\sum_{\nu k} C_{\mu\nu}^k
\left[ \gamma_k 
\int^t dt'  e^{-\gamma_k(t-t')}(|a_\nu(t')|^2-|a_{\nu0}|^2) \right]
a_\mu+f_\mu.
\label{eq:appendTD}
\end{align}
Following the approach of  \secref{spectrum}.A, we linearize \eqref{appendTD} by expanding  the mode amplitudes $a_\mu$ around their steady-state values: $a_\mu= (a_{\mu0}+\delta_\mu)e^{i\phi_\mu}$ (where $\delta_\mu\ll a_{\mu0}$), and  we omit the  terms $\mathcal{O}(\delta_\mu^2)$. Then, we introduce additional variables $\xi_\mu^k$
\begin{equation}
\xi_{\mu}^k=
\gamma_k 
\int^t dt'  e^{-\gamma_k(t-t')}\delta_\mu(t'),
\end{equation}
where $\xi_\mu^k$ is the  time-averaged amplitude deviation of mode $\mu=1\hdots M$ from steady state at the spatial point $k=1\hdots N$. Having introduced the auxiliary variables $\xi_{\mu}^k$, the set of integro-differential equations \eqref{appendTD} turns into a linear system of ODEs, which we solve by  applying several linear-algebraic  transformations to obtain a compact expression for the covariance matrix, as described in detail below.

Introducing the vector $\Phi_\mu\equiv a_{\mu0}\phi_\mu$ , the linear system of ODEs is conveniently written as
\begin{align}
\dot{\delta}_{\mu}&=
-\sum_{\nu k}(2a_{\mu0}a_{\nu0}\mbox{Re}[C_{\mu\nu}^k])
\xi_\nu^k+f_\mu^\mathrm{R},
\label{eq:deltaAppend}\\
\dot{\Phi}_{\mu}&=
-\sum_{\nu k}(2a_{\mu0}a_{\nu0}\mbox{Im}[C_{\mu\nu}^k])
\xi_\nu^k+f_\mu^\mathrm{I}\\
\dot{\xi}_{\mu}^k&=
-\gamma_k \xi_\mu^k+ \gamma_k\delta_\mu.
\label{eq:xiAppend}
\end{align}
To simplify the notation further, we introduce the  $M\times M$ matrices $\mat{A}^k$ and $\mat{B}^k$ ($k=1\hdots N$), with entries
\begin{align}
A_{\mu\nu}^k=
2a_{\mu0}a_{\nu0}\mbox{Re}[C_{\mu\nu}^k])\\
B_{\mu\nu}^k=
2a_{\mu0}a_{\nu0}\mbox{Im}[C_{\mu\nu}^k]),
\end{align}
and we rearrange the set of equations \eqssref{deltaAppend}{xiAppend}  in  a matrix form  [compare with  \eqsref{delta}{phi}]:
\begin{align}
\frac{d}{dt}\boldsymbol{\delta}&=
-\sum_{k} \mat{A}_k \boldsymbol{\xi}\hspace{2pt}^k
+\vec{f}\hspace{2pt}^\mathrm{R},
\label{eq:DeltaAppend}\\
\frac{d}{dt}\boldsymbol{\Phi}&=
-\sum_{k} \mat{B}_k \boldsymbol{\xi}\hspace{2pt}^k
+\vec{f}\hspace{2pt}^\mathrm{I},
\label{eq:PhiAppend}\\
\frac{d}{dt}{\boldsymbol{\xi}}\hspace{2pt}^k&=
-\gamma_k {\boldsymbol{\xi}}\hspace{2pt}^k+ \gamma_k\boldsymbol{\delta}.
\label{eq:XiAppend2}
\end{align}

The autocorrelation matrix of the phase vector $\vec{\Phi}$, which we calculate in this section, is determined by the autocorrelation matrix of the Langevin force
\begin{equation}
\left<\vec{f}(t)\vec{f}^{*T}\hspace{-2pt}(t')\right>=\mat{R}\delta(t-t').
\label{eq:matFDT}
\end{equation}
In order to compute $\left<\vec{\Phi}\vec{\Phi}^T\right>$, we solve \eqref{PhiAppend} by straightforward integration. We find that the phase covariance matrix is a sum of a ``pure'' phase-diffusion term, proportional to $\frac{\mat{R}}{2}$, and  an amplitude--phase coupling term, proportional to $\mathcal{J}$:
\begin{equation}
\left<\vec{\Phi}(t)\vec{\Phi}\hspace{2pt}^T(0)\right>=\left(\frac{\mat{R}}{2}+\mathcal{J}\right)|t|,
\end{equation}
where we have the introduced the shorthand notation
\begin{equation}
\mathcal{J}\equiv 
\frac{1}{|t|}
\sum_{kl}
\mat{B}_k \int\!\!\!\int dt'ds' 
\left<\vec{\xi}_k(t')\vec{\xi}\hspace{2pt}_l^{T}(s')\right>
\mat{B}_l^{T}
\label{eq:subME}
\end{equation}
for the second term, which is responsible for the generalized $\alpha$ factor.

In the remainder of this section, we calculate $\mathcal{J}$. First, we solve the set of ODEs for $\boldsymbol{\xi}^k$ and $\boldsymbol{\delta}$ \eqsref{DeltaAppend}{XiAppend2}, and then we substitute the solution for $\boldsymbol{\delta}$ into \eqref{subME} and evaluate the integrals. To this end, we begin by rewriting  the equations for $\boldsymbol{\xi}^k$ and $\boldsymbol{\delta}$ more compactly. We define the $[(N+1)\cdot M]\times1$ vectors $\vec{x}$ and $\vec{F}$ and the $[(N+1)\cdot M]\times[(N+1)\cdot M]$ matrix $\mat{K}$:
\begin{align}
\vec{x}=
\left( \begin{array}{c}
\boldsymbol{\delta} \\
\boldsymbol{\xi}^1 \\
\vdots
\\
\boldsymbol{\xi}^N \end{array} \right)
\hspace{0.20in}
,
\hspace{0.2in}
\vec{F}=
\left( \begin{array}{c}
\vec{f}^\mathrm{R} \\
\vec{0} \\
\vdots
\\
\vec{0} \end{array} \right)
\nonumber\\
\mat{K}=
\left( \begin{array}{ccccc}
0 & \mat{A}_1 & \mat{A}_2 & \hdots &\mat{A}_N\\
\mat{\Lambda}_1& -\mat{\Lambda}_1 & 0 & \hdots &0\\
\mat{\Lambda}_2 & 0 & -\mat{\Lambda}_2 &   &0\\
\vdots&\vdots&0&\ddots&0
\\
\mat{\Lambda}_N &0&\hdots&0& -\mat{\Lambda}_N \end{array} \right),
\end{align}
where $\mat{\Lambda}_k$ are block-diagonal $M\times M$ matrices with $\gamma_k$ on the diagonal entries, and the zeros in the definition of $\mat{K}$ are block $M\times M$ zero matrices. Using these definitions, the equations for $\boldsymbol{\xi}^k$ and $\boldsymbol{\delta}$ \eqsref{DeltaAppend}{XiAppend2} can be conveniently written as
\begin{equation}
\frac{d}{dt}\vec{x} = -\mat{K}\vec{x}+\vec{F}.
\label{eq:linEq}
\end{equation}
The solution of \eqref{linEq} is
\begin{equation}
x_m(t) = \int^t dt' \sum_\rho \left[
e^{-\mat{K}(t-t')}
\right]_{m\rho}F_\rho(t')
\end{equation}
and, in particular, the solution for $\xi_\mu^k$ is
\begin{equation}
\xi_\mu^k = \int^t dt' \sum_{s=1}^M
\left[
e^{-\mat{K}(t-t')}
\right]_{Mk+\mu,s}
f^\mathrm{R}_\mathrm{S}(t').
\end{equation}
For ease of notation, let us denote the ${(k+1)}^{\mbox{st}}$ $M\times M$ block in the first column of the matrix  $e^{-\mat{K}(t-t')}$ by the shorthand notation $[e^{-\mat{K}(t-t')}]_{k+1,1}$, so that $\boldsymbol{\xi}_k = \int_0^t dt' [e^{-\mat{K}(t-t')}]_{k+1,1}
\vec{f}^\mathrm{R}(t')$. 
Substituting the expression  for $\boldsymbol{\xi}_k$ into $\mathcal{J}$
and using the autocorrelation function of the Langevin force \eqref{matFDT}, we obtain
\begin{align}
\mathcal{J}=\frac{1}{|t|}
\sum_{k\ell}\mat{B}_k \int\!\!\!\int\!\!\!\int dt'dt''ds' 
\left[e^{-\mat{K}(t'-t'')}\right]_{k+1,1}\nonumber\\
\times\hspace{2pt} \frac{\mat{R}}{2} \hspace{2pt}
\left[e^{-\mat{K}^T(s'-t'')}\right]_{\ell+1,1}
\mat{B}_\ell^T.
\label{eq:stepJ}
\end{align}
We proceed (not shown) by diagonalizing the matrix $\mat{K}$ and evaluating the integrals in \eqref{stepJ}. (The intermediate steps depend on the eigenvalues of $\mat{K}$ and the matrix of eigenvectors, but the final result can be expressed in terms of the matrix inverse $\mat{K}^{-1}$). In the long-time limit, we keep the leading order term (which grows linearly in time)  and we obtain
\begin{equation}
\mathcal{J}=
\left(\sum_k \mat{B}_k \left[\mat{K}\right]_{k+1,1}^{-1}\right)
\hspace{2pt} \frac{\mat{R}}{2} \hspace{2pt}
\left(\sum_\ell (\mat{B}_\ell \left[\mat{K}\right]_{\ell+1,1}^{-1})^T\right).
\label{eq:almostJ}
\end{equation}

In order to complete the derivation of the linewidth formula, we use the following identity:
\begin{equation}
\left[\mat{K}\right]_{k+1,1}^{-1} =
\left( \sum_j \mat{A}_j \right)^{-1},
\label{eq:identity}
\end{equation}
which we prove below. Noting that $A_{\mu\nu}=\sum_k A_{\mu\nu}^k$ and $B_{\mu\nu}=\sum_k B_{\mu\nu}^k$ and using the identity \eqref{identity}, we find that \eqref{almostJ} reduces to
\begin{equation}
\mathcal{J}=
\mat{B} \mat{A}^{-1}
\hspace{2pt} \frac{\mat{R}}{2} \hspace{2pt}
\left( \mat{B}\mat{A}^{-1}\right)^T,
\end{equation}
which completes the derivation of the linewidth formula in the most general time-delayed model.  In particular, and somewhat remarkably, the $\gamma$ terms completely cancel in the computation of the first column of the matrix inverse, and drop out of the final result.

\uline{Proof of the identity \eqref{identity}}: We use Schur complement~\cite{Bultheel1997} for the lower-left corner of a matrix inverse:
\begin{equation*}
\left( \begin{array}{cc}
\mat{A} & \mat{B} \\
\mat{C} & \mat{D} \end{array} \right)^{-1}=
\left( \begin{array}{cc}
* & * \\
-\mat{D}^{-1}\mat{C}(\mat{A}-\mat{B}\mat{D}^{-1}\mat{C})^{-1} & * \end{array} \right),
\end{equation*}
($\mat{A}$ and $\mat{D}$ need to be square matrices). Decomposing the matrix $\mat{K}$ into the blocks
\begin{equation*}
\mat{A}=
\left( \begin{array}{c}
0  \end{array} \right)
\hspace{0.125in}
,
\hspace{0.125in}
\mat{B}=
\left( \begin{array}{cccc}
\mat{A}_1 & \mat{A}_2 & \hdots &\mat{A}_N \end{array}\right)
\hspace{0.125in},
\end{equation*}
\begin{equation*}
\mat{C}=
\left( \begin{array}{c}
\mat{\Lambda}_1\\
\mat{\Lambda}_2\\
\vdots\\
\mat{\Lambda}_N  \end{array} \right)
\hspace{0.125in}
'
\hspace{0.125in}
\mat{D}=
\left( \begin{array}{cccc}
 -\mat{\Lambda}_1 & 0 & \hdots &0\\
 0 & -\mat{\Lambda}_2 &   &0\\
\vdots&0&\ddots&0\\
0&\hdots&0& -\mat{\Lambda}_N \end{array} \right),
\end{equation*}
we can  calculate the lower-left corner of $[\mat{K}]_{k+1,1}^{-1}$:
\begin{equation*}
-\mat{D}^{-1}\mat{C}(\mat{A}-\mat{B}\mat{D}^{-1}\mat{C})^{-1} =
(\mat{D}^{-1}\mat{C})[\mat{B}(\mat{D}^{-1}\mat{C})]^{-1}=
\end{equation*}
\begin{equation*}
-\left( \begin{array}{c}
1\\
1\\
\vdots\\
1  \end{array} \right)
\left[
\left( \begin{array}{cccc}
\mat{A}_1 & \mat{A}_2 & \hdots &\mat{A}_N \end{array}\right)
\left( \begin{array}{c}
1\\
1\\
\vdots\\
1  \end{array} \right)
\right]^{-1}=
\end{equation*}
\begin{equation*}
-\left( \begin{array}{c}
1\\
1\\
\vdots\\
1  \end{array} \right)
\left[\sum_\mathrm{I} \mat{A}_\mathrm{I}\right]^{-1}.
\end{equation*}
Therefore, we obtain $\left[\mat{K}\right]_{k+1,1}^{-1} =\left( \sum_j \mat{A}_j \right)^{-1}$.

\appendixTRY{Lemma from Sec. VI.C }{$[\mat{B}\mat{A}^{-1}]_{ij}\propto \frac{a_{i0}}{a_{j0}}$}{{appendixC}}
In \secref{applications}.B, we present a multimode Schawlow--Townes relation, which states that  the linewidths are  proportional to a sum of inverse output powers of all the other modes. This result arises from a  lemma which we prove  here.  We use the standard matrix-inverse formula~\cite{Arfken2006}
\begin{equation}
\mat{A}^{-1}=\frac{1}{\det\mat{A}}\hspace{2pt}\mbox{adj}\mat{A},
\label{eq:matINV}
\end{equation}
where the adjugate matrix is defined as 
\begin{equation}
\mbox{adj}\mat{A}=\left((-1)^{i+j}M_{ij}\right)^T.
\end{equation}
$\mat{M}$ is the cofactor matrix, i.e. the matrix whose $(i,j)$ entry is the determinant of the $(i,j)$ minor of $\mat{A}$ (which is the matrix obtained from $\mat{A}$ by deleting the $i$'th row and the $j$'th column). From the definition of $\mat{A}$ (i.e., $A_{ij}\equiv\mbox{Re}[C_{ij}]a_{i0}a_{j0}$), it follows that
\begin{equation}
M_{ij}=\left(\prod_{k\neq i,j} a_{k0}^2\right) a_{i0}a_{j0} Q_{ij}.
\label{eq:cofactor}
\end{equation}
where $Q_{ij}$ (and later  $Q$)  denote constants that may depend on $i$ and $j$, but are independent of the modal amplitudes. Note also that
\begin{equation}
\det\mat{A}=\prod_{k} a_{k0}^2 \cdot Q.
\label{eq:determinant}
\end{equation}
Using  \eqssref{matINV}{determinant}, we obtain
\begin{equation}
A^{-1}_{ij}=\frac{1}{a_{i0}a_{j0}}\cdot Q_{ij}.
\end{equation}
Therefore, one can easily see that the lemma follows, since $[\mat{B}\mat{A}^{-1}]_{ij}\propto \sum_k a_{i0}a_{k0}\cdot\frac{1}{a_{k0}a_{j0}}\propto\frac{a_{i0}}{a_{j0}}$.

\appendixsec{Comparison with the scattering-matrix linewidth formula}{appendixD}

In a recent scattering-matrix based linewidth theory~\cite{Pillay2014}, Pillay \emph{et al.}~obtain a formula for the linewidth of a one-dimensional  laser system, expressed in terms of integrals over the modes which solve the nonlinear SALT equations.  In this appendix, we prove that their formula (which applies to 1d systems) is equivalent to our linewidth formula \eqref{TCMTwidth} (except that their formula gives a spatially averaged incomplete-inversion factor and omits the $\alpha$ factor). 

In the scattering-matrix approach, the lasing modes are described as purely outgoing wave functions ${\psi}_0$, which satisfy the nonlinear SALT equation
\begin{equation}
\nabla\times\nabla\times {\psi}_0(x)-
\omega_0^2 \varepsilon(x,\omega_0){\psi}_0(x)=0,
\label{eq:YidongSALT}
\end{equation}
and can be expressed as a superposition of outgoing channel modes ${\vec{u}_\mu}$ outside of the laser region 
\begin{equation}
{\psi}_0({x})=\sum_k b_k \vec{u}_k(x,\omega_0)\hspace{0.5in} \mbox{for r}\notin C.
\end{equation}
$C$ denotes the scattering region (i.e., $\varepsilon=1$ for $x\notin C$). Note that ${\psi}_0({x})$ is  precisely the same the mode $\vec{E}_0(\vec{x})$ (which was  used in  \secref{formula}) inside the cavity region.] The outgoing mode-amplitudes $\vec{b}$ are normalized to the value of $\psi_0$ at the cavity boundary ($x=L$)
\begin{equation}
\vec{b}^T\vec{b} = \vec{\psi}_0^2(L).
\label{eq:needShow}
\end{equation}

The apparent difference between our formula and the linewidth formula in~\cite{Pillay2014} is that the integral term in the denominator of our linewidth formula\eqref{TCMTwidth} is replaced by a sum of two terms in the scattering-matrix approach
\begin{align}
\int_{\substack{\text{all}\\\text{space}}}\!\!
dx \left[\varepsilon\omega_0+\frac{\omega_0^2}{2}\frac{d\varepsilon}{d\omega_0}\right]\psi_0^2
\longrightarrow\hspace{1in}
\nonumber\\
\frac{i\vec{b}^T\vec{b}}{2}+\!\int_\mathrm{C}\! dx \left[\varepsilon\omega_0+\frac{\omega_0^2}{2}\frac{d\varepsilon}{d\omega_0}\right]\psi_0^2.
\end{align}
In order for the two formulas to agree, we need to show that
\begin{equation}
\omega_0 \int_L^{\infty}\!\!\!
dx \psi_0^2(x)
=\frac{i\vec{b}^T\vec{b}}{2}
\label{eq:scatteringProof}
\end{equation}
(where we have used the fact that $\varepsilon=1$ outside the cavity region). We show that the latter condition \eqref{scatteringProof} holds for any solution $\psi_0$  of \eqref{YidongSALT} which satisfies outgoing boundary conditions. One way to impose outgoing boundary conditions is to invoke the limiting-absorption principle (i.e., add loss to eliminate incoming waves from infinity and take the limit of infinitesimal absorption at the end of the calculation~\cite{Schulenberger1971,Taflove2013}). Formally, we define the integral on the left-hand side of \eqref{scatteringProof} as
\begin{equation}
\int_L^{\infty}\!\!\!
dx \psi_0^2(x)
\equiv
\lim_{s\rightarrow0^+}
\int_L^{\infty}\!\!\!
dx e^{-sx} \psi_0^2(x).
\label{eq:intDEF}
\end{equation}
By substituting $\psi_0(x)=e^{ik_0x}$ into \eqref{intDEF} and taking the limit of $s\rightarrow0^+$, we obtain 
$\int_L^{\infty}\!\!\!dx \psi_0^2(x)=\frac{i}{2k}e^{2ik_0L}=\frac{i}{2k}\vec{b}^T\vec{b}$, and since $\omega_0=ck_0$ this finishes the proof of \eqref{needShow} (with the units convention of $c=1$).

\appendixsec{Zero-point Fluctuation Cancellation}{appendixE}

The hyperbolic cotangent factor in the FDT \eqref{FDT} arises as a sum of a Bose--Einstein distribution and a $\sfrac{1}{2}$ factor stemming from quantum zero-point (ZP) fluctuations~\cite{Landau1980,Kittel1980}, and this is why it does not vanish in the limit of zero temperature ($\beta \to \infty$).    However, it turns out that contribution of this ZP term cancels in the linewidth formula, as was shown by Henry and Kazarinov~\cite{Henry1996} from a quantum-operator viewpoint, and it is convenient to explicitly subtract the ZP term from the hyperbolic cotangent as in \eqref{TCMTwidth} and \eqref{auto}. Here, we provide a purely classical explanation for why this cancellation occurs, and why it is important to perform the explicit subtraction in order to eliminate a subtlety arising from the definition of outgoing boundary conditions.

The FDT has a hyperbolic cotangent factor,  and when we apply the FDT to find the $\left<f_\mu f_\nu^{*}\right>$ correlation function in Sec IV, the same hyperbolic cotangent factor arises in the $R$ integral, appearing in the form
\begin{align}
\int dx |\vec{E}_\mu|^2 \mbox{Im}\hspace{2pt}\varepsilon_0(\omega_\mu) \cdot\frac{1}{2}\coth\left(\frac{\hbar\omega_\mu\beta}{2}\right) \hspace{1in}\nonumber\\
= \int dx |\vec{E}_\mu|^2 \mbox{Im}\hspace{2pt}\varepsilon_0(\omega_\mu) \left[\left(\frac{1}{2}\coth\frac{\hbar\omega_\mu\beta}{2}-\frac{1}{2}\right)
+\frac{1}{2}\right]
\end{align}
for a lasing mode $\mu$, where we have trivially added and subtracted the ZP $\sfrac{1}{2}$ factor from coth.   Now, we wish to analyze the final $\sfrac{1}{2}$ term, which is the integral $\frac{1}{2}\int dx |\vec{E}_\mu|^2 \mbox{Im}\hspace{2pt}\varepsilon_0(\omega_\mu)$.   Before we treat outgoing boundary conditions, let us consider the simpler case of a laser surrounded by an explicit absorbing medium, as in~\cite{Henry1996}.  (This is also the situation  in more recent computational models, for which one uses a finite spatial domain surrounded by absorbing layers~\cite{Esterhazy2013}.)    For any steady-state lasing mode (real $\omega_\mu$), the net gain + loss is zero, but $\frac{1}{2}\int dx |\vec{E}_\mu|^2 \mbox{Im}\hspace{2pt}\varepsilon_0(\omega_\mu)$ is proportional to the net power absorbed or gained by the electric field~\cite{Jackson1999} and hence this integral is \emph{zero}.   Therefore, in such a case, whether or not we include the $\sfrac{1}{2}$ factor is irrelevant, because the $\pm\sfrac{1}{2}$ terms integrate to zero.

However, a subtlety arises in this integral in the common case where the laser is surrounded by an infinite zero-temperature ($\beta = \infty$) lossless medium with outgoing radiation boundary conditions.   Outgoing boundary conditions can be defined mathematically by the limiting absorption principle~\cite{Schulenberger1971,Taflove2013}: one takes the lossless medium to be the limit of a lossy medium as the losses go to zero from above, which can be expressed by writing $\varepsilon$ as  $\varepsilon + i0^+$.   Just as in appendix D, the correct approach is to take the lossless limit \emph{after} solving the problem, i.e. the $0^+$ limit is taken \emph{outside} of the integral.   Before we take this limit, it makes no difference whether the 1/2 factor is included, just as above: it integrates to zero.   However, after we take the lossless limit, there is no explicit absorbing region ($\mbox{Im}\hspace{2pt}\varepsilon> 0$) in the integral (the absorption has been ``moved to infinity'' in some sense), so if we perform the coth integral without subtracting $\sfrac{1}{2}$ then we would obtain an incorrect contribution from the ZP fluctuations in the gain medium (which should have been canceled).   Instead, if we integrate against $\sfrac{1}{2}\coth -\sfrac{1}{2}$, the result is correct without requiring any explicit contribution from the absorbing boundary conditions.

Note that if the laser is surrounded by an infinite lossless medium at a \emph{positive} temperature, then there is a nonzero contribution of incoming thermal radiation to the linewidth~\cite{Haken1984,Haken1985}.   This can be included in one of two ways.  In practice, we typical solve the SALT equations in a finite computational box with an explicit absorbing region, in which case no modification to our linewidth formula is required: one simply assigns the ambient temperature to the absorbing region.   If, on the other hand, the outgoing boundary conditions are imposed in some other way (e.g. semi-analytically as in earlier SALT work~\cite{Tureci2006,Tureci2007,Tureci2008,Ge2010}), then an explicit source term must be added to account for incoming thermal radiation, as in previous works~\cite{Andreasen2008}.

\bibliography{bibliography}

\end{document}